\documentclass[twocolumn]{autart}    % Enable this line and disable the 
\usepackage{mathtools}
\usepackage{amsmath}
\usepackage{amssymb}
\usepackage{mathrsfs}
\usepackage{graphicx}
\usepackage{subfigure}
\usepackage{epstopdf}
\usepackage{color}

\usepackage{etoolbox}

\let\classAND\AND
\let\AND\relax

\usepackage{algorithm}
\usepackage{algorithmic}

\let\AND\classAND

\AtBeginEnvironment{algorithmic}{\let\AND\algoAND}
\usepackage{cite}

\allowdisplaybreaks[4]
\newcommand{\bfl}{\mathbf{1}}

\newcommand{\EE}{\mathbb{E}}
\newcommand{\PP}{\mathbb{P}}

\begin{document}

\begin{frontmatter}

\title{{\color{black}Community Structure Recovery\\ and Interaction Probability Estimation\\ for Gossip Opinion Dynamics}} % Title, preferably not more 

% \thanks[footnoteinfo]{This paper was not presented at any IFAC 
% meeting. Corresponding author }

% \thanks[footnoteinfo]{Corresponding author }

\author[kth]{Yu Xing}\ead{yuxing2@kth.se},    % Add the 
\author[kth,cor1]{Xingkang He}\ead{xingkang@kth.se}, 
\corauth[cor1]{Corresponding author.}
\author[amss]{Haitao Fang}\ead{htfang@iss.ac.cn}, 
\author[kth]{Karl H. Johansson}\ead{kallej@kth.se}         % e-mail address 
 % (ead) as shown

\address[kth]{Division of Decision and Control Systems, School of Electrical Engineering and Computer Science,\\ KTH Royal Institute of Technology, and Digital Futures, Stockholm, Sweden.}  % Please supply                                              
\address[amss]{Key Laboratory of Systems and Control, Academy of Mathematics and Systems Science, Chinese Academy of Sciences;\\ School of Mathematical Sciences, University of Chinese Academy of Sciences, Beijing, P. R. China.}             % full addresses
% \address[Baiae]{The White House, Baiae}        % here.

                             % keyword list or with the 
                                          % help of the Automatica 
                                          % keyword wizard

\begin{abstract}                          % Abstract of not more than 200 words.
We {\color{black}study how to jointly recover the community structure and estimate the interaction probabilities} of gossip opinion dynamics. In this process, agents randomly interact pairwise, and there are stubborn agents never changing their states. Such a model illustrates how disagreement and opinion fluctuation arise in a social network. It is assumed that each agent is assigned with one of two community labels, and the agents interact with probabilities depending on their labels. {\color{black}The considered problem is to jointly recover the community labels of the agents and estimate interaction probabilities between the agents}, based on a single trajectory of the model. We first study stability and limit theorems of the model, and then propose a joint {\color{black}recovery} and estimation algorithm based on a trajectory. It is verified that the community {\color{black}recovery can be achieved} in finite time, and the interaction estimator converges almost surely. We derive a sample-complexity result for the {\color{black}recovery}, and analyze the estimator's convergence rate. Simulations are presented for illustration of the performance of the proposed algorithm.
\end{abstract}

\end{frontmatter}

\section{Introduction}

Networks %, consisting of nodes and edges, 
appear across domains from biology %, computer science, 
to sociology. Real networks often exhibit community structures, where subsets
of nodes have dense connections locally but sparse connections
globally
\cite{fortunato2016community}. Community detection is to partition nodes according to the network topology. There is a growing interest in studying community detection based on state
observations of dynamics %, without topological data 
\cite{prokhorenkova2022less, peixoto2019network, wai2019blind, schaub2020blind, roddenberry2020exact, ramezani2018community}. 
Lacking topology data makes the problem harder than classic ones. {\color{black}Particularly, it is
unclear how to recover communities out of a single trajectory of opinion dynamics~\cite{ravazzi2021learning}.}

\vspace{-0.2ex}
\subsection{Related work}\label{sec_sub_relatedwork}

%This subsection discusses existing researches related to the considered problem. 
In this subsection, we first review key community definitions and detection approaches~\cite{fortunato2016community}%,schaub2017many}
, then discuss recovering communities based on state observations, and finally clarify our motivation.

Traditional community detection methods apply %agglomerative or divisive 
classic clustering techniques to node pairs assigned with certain weights~\cite{girvan2002community}. %(e.g., the number of node-independent paths\cite{wasserman1994social} or edge betweenness \cite{girvan2002community}). 
In \cite{newman2004finding}, the authors introduce the concept of modularity to measure the quality of a graph partition. 
%High modularity of a graph partition indicates dense connections within communities given by this partition. 
A famous algorithm based on optimizing modularity is the Louvain method \cite{blondel2008fast}, which assigns nodes to one of the communities iteratively to achieve the largest modularity gain.
Another approach to community detection is based on statistical inference%. Instead of defining communities resorting to quality functions such as modularity, this approach
, which introduces generative network models and considers an observed network as a sample. A  canonical model is the stochastic block model (SBM). % %%\cite{%holland1983stochastic,decelle2011asymptotic
%abbe2017community}. 
%, which generates a network by first assigning each node a community label, and then connecting nodes with probabilities depending on their labels. 
%The statistical framework makes it possible to analyze community detectability and performance of detection algorithms theoretically. 
The paper~\cite{abbe2017community} reviews results %on detectability of communities and consistency of algorithms. 
on detectability of the SBM and performance of algorithms.
% (e.g., spectral methods \cite{mcsherry2001spectral} and belief propagation \cite{mossel2016belief}).
Besides optimization and statistical approaches, another method is based on dynamical processes (e.g.,~\cite{rosvall2008maps}). %, are based on dynamical processes over networks. %, for example, the Infomap method~\cite{rosvall2008maps}. 
%The intuition is that %the community structure of a network can somehow influence the behavior of dynamics on the network, so 
%the behavior of dynamics on a network may capture the structure. %, probably higher-order ones beyond edges and node degrees. Random walk is a commonly used dynamical process. 
%For example, the Infomap method~\cite{rosvall2008maps} seeks two-level (coarse and fine) coding of a trajectory of a random walk over a network, and correspond the obtained coarse-level codewords to communities. 
%There are also algorithms based on opinion dynamics models. 
The paper~\cite{morarescu2010opinion} proposes a bounded-confidence model, where agents converge to
several clusters corresponding to communities.

Recently, the study of community detection for networked dynamics has emerged. The problem is to recover communities only based on state observations of a dynamical process.
%, without using topological data. 
The main difference between this problem and the classic ones, especially the dynamic-based methods, is that the network is not available. The papers \cite{prokhorenkova2022less,ramezani2018community} apply maximum likelihood methods to cascade data. The paper \cite{prokhorenkova2022less} also proposes a two-step procedure, first constructing a network and then clustering agents based on the network.
%However, it may be difficult to estimate the underlying network with desired accuracy, resulting in degraded detection performance, if there is insufficient excitation of the system. 
The authors of \cite{wai2019blind,schaub2020blind,roddenberry2020exact} introduce the blind community detection method, using sample covariance matrices of agent states for recovery. The author in~\cite{peixoto2019network} investigates simultaneously reconstructing the topology and the community structure for epidemics and the Ising model. %, but studies the performance of proposed methods via simulation on datasets.
The paper \cite{berthet2019exact} studies recovery for an Ising blockmodel.
%{\color{black}Recovering communities based on observations from an Ising blockmodel is studied in~\cite{berthet2019exact}.}

We study {\color{black}how to jointly recover the community structure and estimate the interaction probabilities of gossip opinion dynamics}. {\color{black}The problem arises from recent investigation of learning interpersonal influence from dynamics \cite{ravazzi2021learning}. }
Network data is useful for decision making, but directly collecting such
data can be hard, due to topic specificity~\cite{cowan2018could}, consistency issues %(e.g., online links may not reflect the ground truth) 
\cite{netrapalli2012learning}, and privacy concern \cite{de2018privacy}.
Learning large-scale networks may be computationally expensive, so {\color{black}recovering} communities as a coarse description is a good option. {\color{black}The gossip update rule %, as a counterpart of the well-known DeGroot model \cite{proskurnikov2017tutorial}, 
captures the random nature of individual interactions.} It is a fundamental element of many opinion models \cite{%deffuant2000mixing,acemouglu2013opinion, ravazzi2014ergodic,
proskurnikov2017tutorial}, and has also been extensively studied % as a consensus algorithm 
\cite{boyd2006randomized}. {\color{black}Stubborn agents, such as media and opinion leaders, play a crucial role in opinion formation \cite{ramos2015does}.} The paper~\cite{acemouglu2013opinion} shows that the existence of stubborn agents can explain opinion oscillation. A generalization of stubborn agents is to assume that each agent has some level of stubbornness with respect to its initial belief. This generalization is considered by the Friedkin–Johnsen model and its extensions \cite{proskurnikov2017opinion,tian2018opinion}. 

\subsection{Contributions}\label{sec_sub_contribution}

We consider {\color{black}jointly recovering communities and estimating interaction probabilities for gossip opinion dynamics}. Each agent is assigned with one of the two community labels, and the agents interact with probabilities depending on their labels. Our contributions are as follows:

1. We study properties of the model by leveraging results on Markov chains and stochastic approximation (SA) (Theorem~\ref{thm_stability}). It is shown that regular-agent states converge in distribution to a unique stationary distribution, and the time average of the agent states converge almost surely. An explicit expression for the mean of the stationary distribution is given (Proposition~\ref{thm_expectation_structure}).\newline
2. {\color{black}We develop a joint algorithm (Algorithm~\ref{alg_1}) to recover the community structure and to estimate the interaction probabilities, based on Polyak averaging and SA techniques. The algorithm is able to recover the communities in finite time, and then able to estimate the interaction probabilities consistently (Theorem~\ref{thm_convergence}).}\newline
%Based on Polyak averaging and stochastic approximation techniques, we propose a joint algorithm to detect the community structure and to estimate the interaction probabilities between agents (Algorithm~\ref{alg_1}). It is shown that the community detector of the algorithm converges in finite time, and that the interaction estimator converges almost surely (Theorem~\ref{thm_convergence}). 
3. {\color{black} We show how to theoretically analyze the developed joint algorithm. A concentration inequality for Markov chains (Lemma~\ref{lem_concentration}) is obtained, and it is used in the sample-complexity analysis of the recovery step (Theorem~\ref{thm_finitesample}).} The obtained result shows that the probability of unsuccessful {\color{black}recovery} decays exponentially over time. Additionally, we analyze convergence rate of the interaction estimator from an SA argument (Theorem~\ref{thm_convergencerate}).

The obtained results indicate that a Polyak averaging technique can be useful for {\color{black}recovering} communities based on a single trajectory% with random interactions
. In addition, we establish a sample-complexity result for successful {\color{black}recovery} ({\color{black}recovering} all community labels correctly), providing a quantitative dependence of the successful {\color{black}recovery} probability on model parameters.
These two points make our paper different from~\cite{wai2019blind, schaub2020blind, roddenberry2020exact}, which use covariance matrices of samples from several trajectories, and different from~\cite{wai2016active}, which considers learning a sparse characterization of the network from the gossip model.
{\color{black}The considered problem is different from classic system identification~(e.g.,~\cite{JMLR:v22:19-725}), because stubborn agents normally have fixed states, which does not satisfy input conditions required for system identification, and also because community recovery cannot be obtained directly from parameter estimates.}
% There are two key differences between our paper and previous studies. First, the paper focuses on community detection for a gossip model by using properties of states directly, rather than utilizing intermediate estimates of the underlying network \cite{prokhorenkova2019learning,ramezani2018community}. Additionally, we consider an online community detection problem, aiming at recovering the community structure gradually as the process evolves. This is different from \cite{wai2019blind, schaub2020blind, roddenberry2020exact}, where the state vector at the same time step needs to be observed in several realizations of the considered dynamics.
The major differences between this paper and its conference version \cite{xing2020community} are that we clarify our assumptions in more detail, characterize the sample complexity and the convergence rate of the algorithm, and add more numerical experiments to illustrate its performance.

% \subsection{Notation and preliminaries}\label{sec_sub_notation} 

\subsection{Outline}\label{sec_sub_outline}
The rest of the paper is organized as follows. Section~\ref{sec_problem} formulates the problem. Analysis of the model is given in Section \ref{sec_model}, and a joint {\color{black}recovery} and estimation algorithm is proposed in Section \ref{sec_alg}. Section \ref{sec_alg_analysis} presents convergence results of the algorithm, and Section \ref{sec_simulation} provides several numerical experiments. Finally, Section \ref{sec_conclusion} concludes the paper. Some proofs are postponed to appendices.

\textbf{Notation.}
Denote the $n$-dimensional Euclidean space by $\mathbb{R}^n$, the set of $n\times m$ real matrices by $\mathbb{R}^{n\times m}$, the set of nonnegative integers by $\mathbb{N}$, and the set of positive integers by $\mathbb{N}^+$. Let $\mathbf{1}_n$ be the all-one vector with dimension $n$, $\textbf{e}_i$ be the unit vector with $i$-th entry being one, $I_n$ be the $n\times n$ identity matrix, and $\mathbf{0}_{n,m}$ be the $n\times m$ all-zero matrix. Define $\mathbf{1}_{n_1,n_2} := \mathbf{1}_{n_1} \mathbf{1}_{n_2}^T$, where $A^T$ represents the transpose of a matrix $A$. Denote the Euclidean norm of a vector by $\|\cdot\|$, and denote the maximum absolute column sum norm, spectral norm, and maximum absolute row sum norm of a square matrix by $\| \cdot \|_1$, $\|\cdot\|$, and $\| \cdot \|_{\infty}$. Denote the diagonal matrix with the elements of a vector $x$ on the main diagonal by $\text{diag}\{x\}$.

For a vector $x\in \mathbb{R}^n$, denote its $i$-th component by $x_i$, and for a matrix $A = [a_{ij}]_{1\le i,j \le n} \in \mathbb{R}^{n\times n}$, denote its $(i,j)$-th entry by $a_{ij}$ or $[A]_{ij}$. The matrix $A$ is said to be row stochastic if $a_{ij}\ge 0$ and $A\mathbf{1} = \mathbf{1}$, and to be substochastic if $a_{ij}\ge 0$ and the row sums of $A$ are not larger than one. Denote the spectral radius of $A$ by $\rho(A)$. The cardinality of a set $\Omega$ is denoted by $|\Omega|$. The function $\mathbb{I}_{[\textup{property}]}$ is the indicator function equal to one if the property in the bracket holds, and equal to zero otherwise. For two sequences $\{a_k\}$ and $\{b_k\}$ with $a_k \in \mathbb{R}^n$ and $0 \not= b_k \in \mathbb{R}$, $k \ge 1$, $a_k = O(b_k)$ means that $\|a_k/b_k\| \le C$ for all $k$ and some $C>0$, and $a_k = o(b_k)$ means that $\lim_{k \to \infty} \|a_k/b_k\| = 0$. An event happens almost surely (a.s.) if it happens with probability one. $\mathbb{E}\{X\}$ is the expectation of the random vector $X$. 
%We use the notations $\overset{\textup{P}}{\to}$ and $\overset{\textup{d}}{\to}$ to denote convergence in probability and in distribution of random vectors, respectively.
The notation $\overset{\textup{d}}{\to}$ represents convergence in distribution. 

To define a Markov chain taking values on $(\mathbb{R}^n, \mathcal{B}(\mathbb{R}^n))$, where $\mathcal{B}(\mathbb{R}^n)$ is the Borel $\sigma$-field, we first define the transition probability kernel $P(x, A)$, $x \in \mathbb{R}^n$, $A \in \mathcal{B}(\mathbb{R}^n)$, satisfying that for each $A \in \mathcal{B}(\mathbb{R}^n)$, $P(\cdot, A)$ is a non-negative measurable function on $\mathbb{R}^n$, and for each $x \in \mathbb{R}^n$, $P(x, \cdot)$ is a probability measure on~$\mathcal{B}(\mathbb{R}^n)$. A (homogeneous) Markov chain $\{X(t), t \in \mathbb{N}\}$ on $\mathbb{R}^n$ satisfies that for all $t\in \mathbb{N}$, $A \in \mathcal{B}(\mathbb{R}^n)$, and $x \in \mathbb{R}^n$,
\begin{align*}
    &\mathbb{P}\{X(t+1) \in A | X(t) = x, X(t-1)\dots, X(0)\} \\
    &= \mathbb{P}\{X(t+1) \in A | X(t) = x\} = P(x, A).
\end{align*}
Using transition probability kernel $P(\cdot,\cdot)$, we can define $n$-step transition probability of $\{X(t)\}$ inductively by
\begin{align*}
    P^t(x,A) = \int_{\mathbb{R}^n} P(x, dy) P^{t-1}(y, A), ~t \in \mathbb{N}^+,
\end{align*}
and $P^0(x,A) = \mathbb{I}_{[x\in A]}$, for all $x\in \mathbb{R}^n$ and $A\in \mathcal{B}(\mathbb{R}^n)$. A stationary distribution of a Markov chain $\{X(t)\}$ with transition probability kernel $P(\cdot, \cdot)$ is a probability measure $\pi$ on $\mathcal{B}(\mathbb{R}^n)$ such that
\begin{align*}
    \pi(A) = \int_{\mathbb{R}^n} \pi(dx) P(x,A), \quad A\in \mathcal{B}(\mathbb{R}^n).
\end{align*}
%Let $\mu$ and $\nu$ be two probability measures on $\mathbb{R}^n$. The Wasserstein metric between $\mu$ and $\nu$ (with respect to Euclidean distance) is
%\begin{align*}
%    d_W(\mu,\nu) = \inf_{(X,Y) \in J} \mathbb{E}\{d(X,Y)\},
%\end{align*}
%where $d(\cdot,\cdot)$ is the Euclidean distance, and $J$ is the set of pairs of random vectors $(X,Y)$ such that the marginal distribution of $X$ is $\mu$ and the marginal distribution of $Y$ is $\nu$. 

% For  $\mu$ and $\nu$ on $\mathcal{B}(\mathcal{X})$, denote the total variation distance by 
% \begin{align*}
%     \|\mu - \nu\|_{\text{TV}} := \sup_{A \in \mathcal{B}(\mathcal{X})} |\mu(A) - \nu(A)|.
% \end{align*}
% A Markov chain with transition probability kernel $P(\cdot, \cdot)$ is called uniformly ergodic if
% \begin{align*}
%     \sup_{x\in \mathcal{X}} \|P^t(x, \cdot) - \pi(\cdot)\|_{\text{TV}} \to 0, \quad t \to \infty.
% \end{align*}

\begin{figure*}
    \centering
    \subfigure[\label{fig_interaction}The left (right) graph demonstrates the case where $w_s > w_d$ ($w_s < w_d$), in which agents within the same community interacting more (less) often than agents between communities. The width of edges is proportional to the number of interactions.]{\quad
    \includegraphics[width=0.23\linewidth]{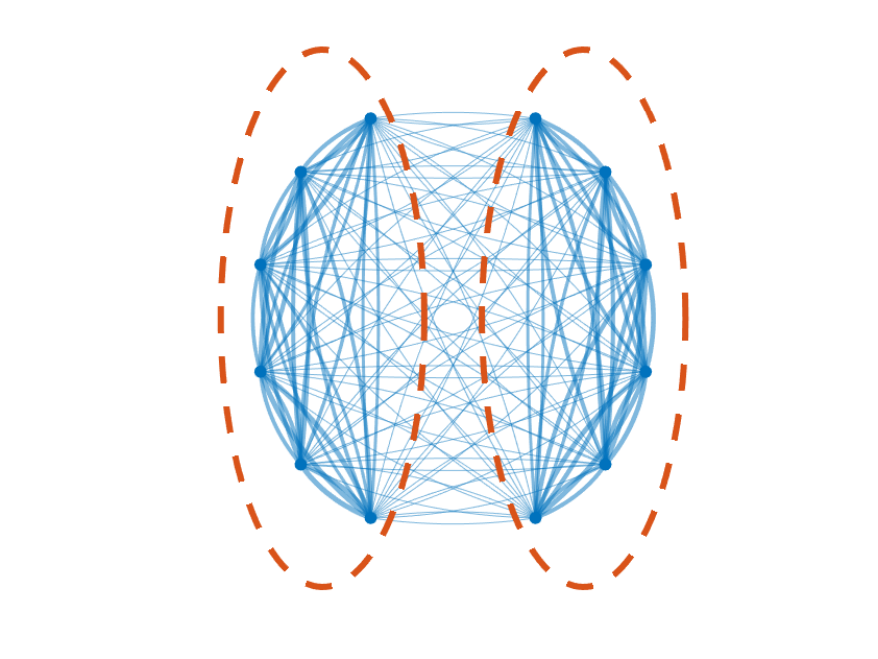}
    \includegraphics[width=0.23\linewidth]{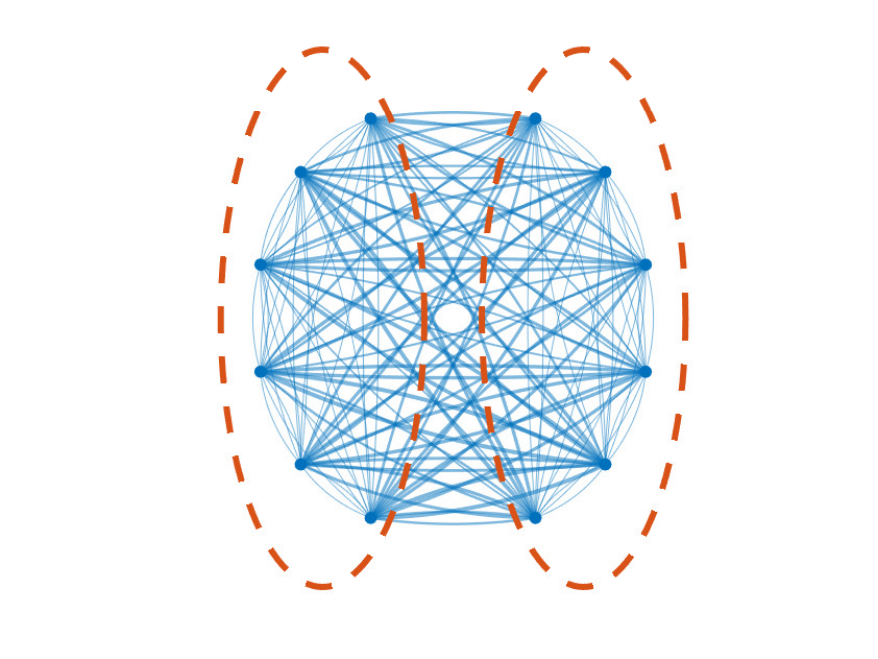}\quad}\quad
    \subfigure[\label{fig_sbm_exmp}The adjacency matrix of a graph generated from $\textup{SBM}(n,\nu_1,\nu_2,p_s,p_d)$ with $n=5000$, $\nu_1 = 0.4$, $\nu_2 = 0.6$, $p_s = 5 \log n/n$, $p_d = \log n/n$. Dots represent nonzero entries, so the block structure of the matrix is clearly visible.]{\qquad\qquad\qquad\quad
    \includegraphics[width=0.18\linewidth]{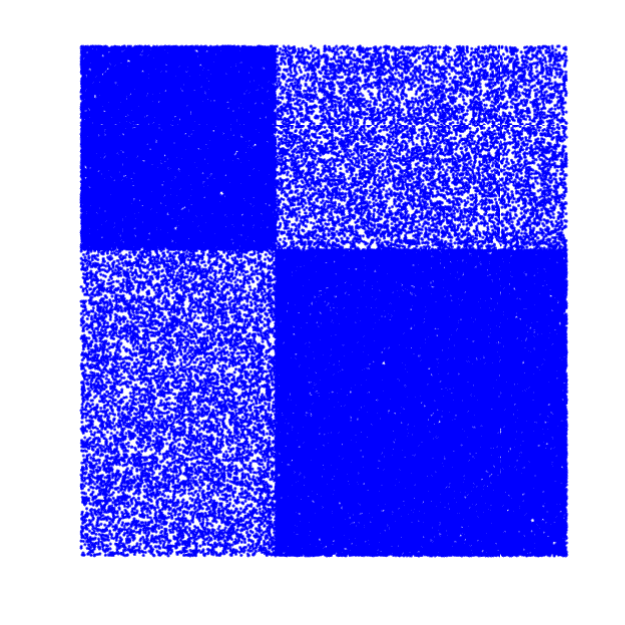} \qquad\qquad\qquad\quad}
    \caption{Illustration of the interaction model \eqref{eq_interaction_prob} and an adjacency matrix generated from an SBM.}    
    %Illustration of agent interaction behavior under different interaction probabilities. The network consists of twelve agents, and the dashed ellipses show different communities. The width of edges is proportional to interaction times between agents.
\end{figure*}

\section{Problem Formulation}\label{sec_problem}

This section introduces the considered model and the definition of communities, and formulates the problem.

\subsection{Gossip Model with Stubborn Agents} 
%Gossip models have been studied extensively in the control literature. They serve as a randomized version of concensus algorithms \cite{??} and a crucial type of update rule in opinion dynamics models \cite{??}. 

%In this section, we present the gossip model.
The gossip model is a random process over an undirected graph $\mathcal{G} = (\mathcal{V}, \mathcal{E})$ with the agent set $\mathcal{V}$, the edge set $\mathcal{E}$, and no self-loops. The agents have two types, regular and stubborn, denoted by $\mathcal{V}_r$ and $\mathcal{V}_s$, respectively ($\mathcal{V} = \mathcal{V}_r \cup \mathcal{V}_s$, $\mathcal{V}_r \cap \mathcal{V}_s = \emptyset$). Each agent $i$ has a state $X_i(t) \in \mathbb{R}$, and the state vector at time $t\in \mathbb{N}$ is $X(t) \in \mathbb{R}^n$. Stubborn agents do not change their states during the process.
 
An interaction probability matrix $W = [w_{ij}] \in \mathbb{R}^{n\times n}$ captures agent interactions, where $w_{ij} = w_{ji} \ge 0$, $w_{ij} > 0 \Leftrightarrow \{i,j\}\in\mathcal{E}$, $i,j\in \mathcal{V}$, and $\mathbf{1}^T W \mathbf{1}/2 =$~$1$. At time $t$, edge $\{i,j\}$ is selected with probability $w_{ij}$ independently of previous updates, and agents update as~follows, {\color{black}with the averaging weight $q \in[0,1)$,} 
\begin{equation}\label{eq_update_rule1}
    X_k(t+1) = \begin{cases}
            q X_k(t) + (1-q) X_l(t), & \text{if } k \in \mathcal{V}_r \cap \{i,j\},\\
            & l \in \{i,j\}\setminus k,\\
            X_k(t), & \text{otherwise}.
            \end{cases}  
\end{equation} 
For $1 \le i < j \le n$, define
\begin{equation*}
	\begin{aligned}
    &R^{ij} = \\
    &\begin{cases}
    I - (1-q) (\textbf{e}_i - \textbf{e}_j)(\textbf{e}_i - \textbf{e}_j)^T, & \text{if } i, j \in \mathcal{V}_r,\\
    I - (1-q) \textbf{e}_i(\textbf{e}_i - \textbf{e}_j)^T, & \text{if } i \in \mathcal{V}_r,  j \in \mathcal{V}_s,\\
    I - (1-q) \textbf{e}_j(\textbf{e}_j - \textbf{e}_i)^T, & \text{if } i \in \mathcal{V}_s,  j \in \mathcal{V}_r,\\
    I, & \text{if } i, j \in \mathcal{V}_s,
    \end{cases}
    \end{aligned}
\end{equation*}
and a sequence of independent and identically distributed (i.i.d.) $n$-dimensional random matrices $\{R(t),$ $t\in \mathbb{N}\}$ such that $ \mathbb{P}\{R(t) = R^{ij}\} = w_{ij}$, $1 \le i < j \le n.$
The update rule \eqref{eq_update_rule1} can then be written as
\begin{align}\label{eq_update_compact}
    X(t+1) = R(t) X(t).
\end{align}
Since stubborn agents never change their states, we rewrite \eqref{eq_update_compact} to end up with the following compact form of \emph{the gossip model with stubborn agents}:
\begin{align}\label{eq_update_compact_regular}
    X^r(t+1) = A(t) X^r(t) + B(t) X^s(t),
\end{align}
where $X^r(t)$ and $X^s(t)$ are the state vectors obtained from stacking the states of regular and stubborn agents, respectively, $X^s(t) \equiv X^s(0)$, and $[A(t) ~ B(t)]$ is the matrix obtained from stacking rows of $R(t)$ corresponding to regular agents. So $\{[A(t) ~ B(t)], t\in \mathbb{N}\}$ is a sequence of i.i.d. random matrices. Assume that the initial vector $X(0)$ is fixed, for simplicity. If $X(0)$ is random, we can study the model by conditioning on  realizations of $X(0)$.

\subsection{Communities} 

We follow the framework of SBMs \cite{abbe2017community} and {\color{black}Ising blockmodels~\cite{berthet2019exact}}, and assume that agents have pre-assigned community labels. We define a community as the set of agents that have the same label. %In the framework of the SBM, the probability that two agents have an edge depends on their labels. This model is simple but can generate various kinds of topological structure.

In particular, we consider the scenario where the network has two disjoint communities, $\mathcal{V}_1$ and $\mathcal{V}_2$% ($\mathcal{V} = \mathcal{V}_1 \cup \mathcal{V}_2$ and $\mathcal{V}_1 \cap \mathcal{V}_2 = \emptyset$)
. Denote the community label of $i$ by $\mathcal{C}(i)$, so $\mathcal{C}(i) = k$ for $i \in \mathcal{V}_k$, $k = 1,2$. We call $\mathcal{C}$ the community structure of the network.
We further assume that the interaction probability of the agents $i$ and~$j$ with $i\not = j$ is
\begin{align}\label{eq_interaction_prob}
    w_{ij} = \begin{cases}
    w_s, & \text{ if } \mathcal{C}(i) = \mathcal{C}(j),\\
    w_d, & \text{ if } \mathcal{C}(i) \not= \mathcal{C}(j),
    \end{cases}
\end{align}
where $w_s, w_d \in (0,1)$ and $w_s \not= w_d$. Thus agents in the same community (different communities) interact with probability $w_s$ ($w_d$). Fig.~\ref{fig_interaction} illustrates two different interaction models, via a simulation where a gossip model defined by~\eqref{eq_interaction_prob} is run for $2000$ iterations and the number of interactions between agents is counted.
To ease notation, we assume that $\mathcal{V}_1 = \{1, \dots, n_1\}$ and $\mathcal{V}_2 = \{n_1 + 1, \dots, n_1 + n_2\}$ with $n_k := |\mathcal{V}_k|$, $k=1,2$, and $n_1 + n_2 = n$. Thus the interaction probability matrix is\vspace{-6ex}
\begin{small}
\begin{equation}\label{eq_block_W}
    W = \begin{bmatrix}
    w_s \mathbf{1}_{n_1,n_1}- \text{diag}\{w_s \mathbf{1}_{n_1}\}& w_d \mathbf{1}_{n_1,n_2}\\
    w_d \mathbf{1}_{n_2,n_1} & w_s \mathbf{1}_{n_2,n_2} - \text{diag}\{w_s \mathbf{1}_{n_2}\}
    \end{bmatrix}
\end{equation}
\end{small}%with $\mathbf{1}^T W \mathbf{1}/2 = 1$ (i.e., only one pair of agents interact at each time). 
It has a block structure corresponding to the community structure of the network. 

The following example illustrates how the preceding assumption arises naturally from an SBM. It shows that
a graph generated from an SBM defines an interaction probability matrix close to an averaged version with the same structure as~\eqref{eq_block_W}.

\begin{exmp}\label{exmp_sbm}
Consider an SBM with two communities, commonly studied in community detection~\cite{abbe2017community}. {\color{black}Such an SBM is a random graph, denoted by  $\textup{SBM}(n,\nu_1,\nu_2,p_s,p_d)$. Here $n$ is the number of agents, $\nu_1 \in (0,1)$ (resp. $\nu_2 \in (0,1)$) is the portion of agents with community label~$1$ (resp. label~$2$), where $\nu_1 + \nu_2 = 1$ and $\nu_1n$ and $\nu_2n$ are integers, and $p_s, p_d \in (0,1)$ are the link probabilities between agents in the same and in different communities. We assume $\mathcal{C}(i) = 1$, $1\le i \le \nu_1n$, and $\mathcal{C}(i) = 2$, $\nu_1n + 1\le i \le n$. 

The $\textup{SBM}(n,\nu_1,\nu_2,p_s,p_d)$ randomly generates an undirected graph $\mathcal{G} = (\mathcal{V},\mathcal{E},\mathcal{A})$: for $i\not= j$, $\{i,j\} \in \mathcal{E}$ with probability $p_s$ if $\mathcal{C}(i) = \mathcal{C}(j)$ and with probability $p_d$ if $\mathcal{C}(i) \not= \mathcal{C}(j)$, independently of other edges.} Here $A = [a_{ij}]$ is the adjacency matrix.
%$l_s$ and $l_d$ are two different positive constants, and $\nu_1$ and $\nu_2$ are positive constants such that both $\nu_1 n$ and $\nu_2 n$ are integers and $\nu_1 + \nu_2 = 1$. $\textup{SBM}(n,l_s,l_d,\nu_1,\nu_2)$ generates an undirected graph $\mathcal{G} = (\mathcal{V},\mathcal{E},\mathcal{A})$, where $\mathcal{A}$ is the adjacency matrix (i.e., $[\mathcal{A}]_{ij} = [\mathcal{A}]_{ji} = \mathbb{I}_{[\{i,j\} \in \mathcal{E}]}$), according the following rule. In the first step, the model assigns $\nu_1 n$ agents with community label~$1$, and $\nu_2n$ agents with community label~$2$. %(we denote the community label of agent~$i$ by $\mathcal{C}(i)$, $1\le i \le n$, and the two communities by $\mathcal{V}_1$ and $\mathcal{V}_2$). 
%In the second step, for all $i,j\in\mathcal{V}$ with $i\not=j$, the model adds edge $\{i,j\}$ to $\mathcal{E}$ with probability $(p_{ij} \ln n)/n$ independently of other edges, where $p_{ij} = l_s$ if $\mathcal{C}(i) = \mathcal{C}(j)$ and $p_{ij} = l_d$ if $\mathcal{C}(i) \not= \mathcal{C}(j)$.
The graph $\mathcal{G}$ defines a gossip model with the interaction matrix $\tilde{W} = \mathcal{A}/\alpha$ and $\alpha = \sum_{i=1}^n\sum_{j=i+1}^n a_{ij} = |\mathcal{E}|$. %We consider the scenario where $n \to \infty$, and t
{\color{black}The inequality 
\[
	 \bigg\| \tilde{W} - \frac{\mathbb{E}\{\mathcal{A}\}}{\mathbb{E}\{\alpha\}} \bigg\| \le \frac{C}{n}
\] holds with a constant $C$, except for a probability vanishing as $n \to \infty$, if $\log n/n = O(\min\{p_s,p_d\})$ (see Appendix~\ref{append_proof_exmp1} for a proof). 
This result implies that, if the network of the gossip model is generated from the SBM, then the interaction probability matrix of the gossip model is close to $\mathbb{E}\{\mathcal{A}\}/\mathbb{E}\{\alpha\}$ when $n$ is large. Note that $\mathbb{E}\{\mathcal{A}\}/\mathbb{E}\{\alpha\}$ has exactly the same structure as $W$ in~\eqref{eq_block_W} with $n_k = \nu_k n$, $k=1,2$, $w_s = p_s/\mathbb{E}\{\alpha\}$, and $w_d = p_d/\mathbb{E}\{\alpha\}$. Fig.~\ref{fig_sbm_exmp} demonstrates this concentration phenomenon with an obvious two-block structure.
The concentration indicates that behavior of the gossip model over a graph generated from the SBM may not deviate too far from the gossip model over the averaged graph, when $n$ is large. In fact, in~\cite{xing2022concentration} we show that the expected stationary states of the two models are close, if $\log n/n = o(\min\{p_s,p_d\})$. This result indicates that the gossip model over the averaged graph can be considered as an approximation of the model over the SBM, and results for the former model can be extended to the latter model.}
\end{exmp}

\begin{rem}
%Note that the community label is unique up to a permutation, by which we mean a redistribution of the labels (e.g., we can call $\mathcal{V}_1$ as “community~2", and $\mathcal{V}_2$ as “community~1"). 
{\color{black}A general assumption for community labels in the SBM is that each agent gets a label~$k$ with probability $\nu_k$ independently of each other, $k=1,2$. This is essentially equivalent to the label assignment with deterministic node portions when $n\to \infty$ (Remark~3 of~\cite{abbe2017community}). Note that it is possible to extend the fixed-label assumption considered in Example~\ref{exmp_sbm} to the deterministic-portion assumption, by conditioning on each assignment and using the law of total probability. The condition $\log n/n = O(\min\{p_s,p_d\})$ implies that the expected agent degree is at least $O(\log n)$. In this case, the SBM generates connected graphs with high probability. The difference between $p_s$ and $p_d$ has to be large enough to make exact recovery possible~\cite{abbe2017community}. Here we consider the dynamics over the averaged graph, so the detectability only requires $w_s\not=w_d$ (Assumption~\ref{asmp_community}~(ii)). Future work will study detectability in the SBM case.} %From the example, we can see that $w_s$ and $w_d$ have another physical meaning: the ratio of the probability of two agents linked by an edge and the expected number of edges in the SBM. %We also note that the case of $w_d > w_s$ corresponds to that of $l_d > l_s$ in the SBM, and it is called the disassortative structure \cite{fortunato2016community}.
\end{rem}

\subsection{Community {\color{black}Recovery} and Interaction Estimation} 
{\color{black}The considered problem is to recover the community structure and to estimate the interaction probabilities} based on state observations, as follows. %The estimate of the interaction probabilities provides a more detailed description of the relationships between agents. %This information could be used for future prediction, as discussed in Section \ref{sec_simul_real}. 
%Formally speaking, we consider the following problem.

\textbf{Problem. } Given a trajectory of the gossip model with the interaction matrix~\eqref{eq_block_W}, develop an algorithm to jointly {\color{black}recover} the community structure $\mathcal{C}$ and estimate the interaction probabilities $w_s$ and $w_d$. 

\begin{rem}\label{rem_problem1}
	{\color{black}  In the preceding problem, we assume that the developed algorithm uses data coming from the gossip model over the averaged graph. A natural question is how this algorithm performs if it uses a trajectory of the gossip model over a graph sampled from an SBM. In Section~\ref{sec_simulation}, we illustrate through simulation that the algorithm performs well also in the SBM case. Such performance is guaranteed by that these two processes behave similarly in terms of their stationary states, as explained in Example~\ref{exmp_sbm}. }
{\color{black}We use ``community recovery'' instead of ``community detection'' to avoid ambiguity, following the terminology of~\cite{berthet2019exact}, because here agent behavior depends directly on the community
structure. %A related problem mentioned in Remark~\ref{rem_problem1} is to consider that the process defined by a graph sampled from the SBM. We will study this problem in the future.
}
\end{rem}

Recall $\mathcal{V}_1 = \{1, \dots, n_1\}$ and $\mathcal{V}_2 = \{n_1 + 1, \dots, n_1 + n_2\}$. 
%Note that we do not know $n_1$ and $n_2$ in advance. Without losing generality, 
We further sort the agents as follows: $\mathcal{V}_{r1} = \{1, \dots, n_{r1}\}$, $\mathcal{V}_{s1} = \{n_{r1} + 1, \dots, n_{1}\}$, $\mathcal{V}_{r2} = \{n_1+1, \dots, n_1 + n_{r2}\}$, and $\mathcal{V}_{s2} = \{n_1 + n_{r2} + 1, \dots, n\}$. Here, $\mathcal{V}_{rk}$ (resp. $\mathcal{V}_{sk}$) is the set of regular (resp. stubborn) agents in the community~$k$, $k=1,2$. Denote $n_{rk} := |\mathcal{V}_{rk}|$, $n_{sk} := |\mathcal{V}_{sk}|$, $n_r := |\mathcal{V}_r|$, and $n_s := |\mathcal{V}_s|$.
%Known? unknown? [$n_1$,$n_2$,$n_{r1}$,...] the structure of asmps (e.g., known stubborn-agent states $\to$ interaction estimation)? reply to reviewers comments 
In the considered problem, the total number of agents is known in advance, the network has two communities, and the stubborn-agent states are observable. But difficulty still remains since $n_k$, $n_{rk}$, $n_{sk}$, $k=1,2$, and interaction information are unknown. The interaction information cannot be obtained in general situations (e.g.,  agent states are only observed at some time steps, or observations are corrupted by noise, as discussed in Remark~\ref{rem_ergodic}).

\section{Model analysis}\label{sec_model}

This section studies model behavior, and provides an explicit expression for the mean of the stationary distribution. Assumptions are summarized as follows. 

\begin{assum}\label{asmp_community}~\\
(i.1) The agent set $\mathcal{V}$ consists of two communities, $\mathcal{V}_1 = \{1, \dots, n_1\}$ and $\mathcal{V}_2 = \{n_1 + 1, \dots, n_1 + n_2\}$ with $n_1, n_2 > 0$ and $n_1 + n_2 = n$. \\
(i.2) Both communities have regular agents, namely, $1 \le n_{r1} \le n_1$, $1 \le n_{r2} \le n_2$.\\
(ii) The interaction probability matrix $W$ has a block structure \eqref{eq_block_W} with $w_s, w_d > 0$, $w_s \not = w_d$, and 
\begin{align}\label{eq_ws_wd}
(n_1(n_1-1) + n_2(n_2-1))w_s + 2n_1n_2 w_d = 2.
\end{align}
(iii) $X(0)$ is deterministic. It holds that $X^r(0) \in \mathcal{S}$ with
\begin{equation}\label{eq_mathcalS}
\mathcal{S} := \{x^r \in \mathbb{R}^{n_r} : x_i^r \in [\underline{s}, \overline{s}], 1\le i \le n_r\},
\end{equation}
where $\underline{s} := \min_{1\le i\le n_s}\{\mathbf{x}_i^s\}$, $\overline{s}:= \max_{1\le i\le n_s} \{\mathbf{x}_i^s\}$, $\mathbf{x}^s := X^s(0) = [(\mathbf{x}^{s1})^T ~ (\mathbf{x}^{s2})^T]^T$ is the stubborn state vector, and $\mathbf{x}^{sk}$ is the vector for the community~$k$, $k=1,2$. 
\end{assum}

\begin{figure*}[ht]
\begin{equation}\label{eq_barA}
%\label{eq_barR}
%    \bar{R} = \frac12
%    \begin{bmatrix}
%    (2 - w_s n_1 - w_d n_2) I_{n_{r1}} + w_s \mathbf{1}_{n_{r1},n_{r1}}~ & w_s \mathbf{1}_{n_{r1},n_{s1}} & w_d \mathbf{1}_{n_{r1},n_{r2}} & w_d \mathbf{1}_{n_{r1},n_{s2}}\\
%    \mathbf{0} & 2I_{n_{s1}} & \mathbf{0} & \mathbf{0}\\
%    w_d \mathbf{1}_{n_{r2},n_{r1}} & w_d \mathbf{1}_{n_{r2},n_{s1}}~ & (2 - w_s n_2 - w_d n_1) I_{n_{r2}} + w_s \mathbf{1}_{n_{r2},n_{r2}}~ & w_s \mathbf{1}_{n_{r2},n_{s2}}\\
%    \mathbf{0} & \mathbf{0} & \mathbf{0} & 2I_{n_{s2}}
%    \end{bmatrix}\\
	\begin{aligned}
    \bar{A} &= I_{n_r} - (1-q)
    \begin{bmatrix}
    a_1 I_{n_{r1}} - w_s \mathbf{1}_{n_{r1},n_{r1}} & -w_d \mathbf{1}_{n_{r1},n_{r2}}\\
    -w_d \mathbf{1}_{n_{r2},n_{r1}} & a_2 I_{n_{r2}} - w_s \mathbf{1}_{n_{r2},n_{r2}}
    \end{bmatrix},\quad
    \bar{B} = (1-q)
    \begin{bmatrix}
    w_s \mathbf{1}_{n_{r1},n_{s1}} & w_d \mathbf{1}_{n_{r1},n_{s2}}\\
    w_d \mathbf{1}_{n_{r2},n_{s1}} & w_s \mathbf{1}_{n_{r2},n_{s2}}
    \end{bmatrix}, \\
    a_k &= w_s n_{k} + w_d n_{3-k},~k=1,2.
	\end{aligned}
\end{equation}
\end{figure*}

\begin{rem}
In Assumption \ref{asmp_community} (i.1), the order of agents is sorted for convenience, but we do not know which group each agent belongs to, before community {\color{black}recovery}. It is necessary to assume $w_s\not= w_d$. Otherwise, $W$ has no block structure. Regular agents are assumed to start from $\mathcal{S}$, which is reasonable and intuitively means that regular states lie between the extreme stubborn states.
%One may recover $A(t)$ and $B(t)$ by finding agents changing their states at each time. But we do not investigate this in detail, because our focus is to detect the community structure by directly using properties of agent states. Moreover, the former method is not valid when observations of states are corrupted by noise, as discussed in Remark~\ref{rem_ergodic}.
\end{rem}

Before studying model behavior, we explicitly write the block structures of 
$\bar{A} := \mathbb{E}\{A(t)\}$ and $\bar{B} := \mathbb{E}\{B(t)\}$ in the following proposition, which says that the block structure of $W$ results in similar agent updates in the same community. 

\begin{prop}\label{thm_barRAB}
Suppose Assumption \ref{asmp_community} holds. Then $\bar{A}$ and $\bar{B}$ have block structures given in~\eqref{eq_barA}.% and~\eqref{eq_barB}, respectively.
\end{prop}

\begin{pf}
For $i < j$, $i, j \in \mathcal{V}_{r}$, $R(t) = R^{ij} = I - (1-q) (\textbf{e}_i - \textbf{e}_j)(\textbf{e}_i - \textbf{e}_j)^T$ with probability $w_{ij}$, so $\bar{r}_{ij} = (1-q) w_{ij}$, where $\bar{r}_{ij}$ (resp. $w_{ij}$) is the $(i,j)$-th entry of $\bar{R}$ (resp. $W$). If~$i$ and~$j$ are in the same community, then $w_{ij} = w_s$. Otherwise, $w_{ij} = w_d$. The values of other off-diagonal entries of $\bar{R}$ can be obtained by following the same argument and the definition of $R^{ij}$. For the diagonal entries of $\bar{R}$, note that $R(t)$ is row stochastic a.s., so $\bar{r}_{ii} = 1 - \sum_{j\not= i} \bar{r}_{ij}$. By comparing $R(t)$ with $A(t)$ and $B(t)$ in the gossip model, we can obtain~\eqref{eq_barA}.% and~\eqref{eq_barB}. 
\hfill$\Box$
\end{pf}

\begin{cor}\label{cor_rho_barA}
If Assumption \ref{asmp_community} holds and there exists at least one stubborn agent in the network (i.e., $n_{r} < n$), then $\bar{A}$ is Schur stable, namely, $\rho(\bar{A}) < 1$.
\end{cor}

\begin{pf}
We know from Proposition \ref{thm_barRAB} that $\bar{A}$ has the form \eqref{eq_barA}. If there exists at least one row of $\bar{A}$ with row sum less than one. So from Lemma \ref{lem_substochastic} in Appendix~\ref{append_proof_thm_stability}, the corollary follows.\hfill$\Box$
\end{pf}

Now we provide the stability and limit theorems of the gossip model.

\begin{thm}(Stability and limit theorems)\label{thm_stability} 
Suppose that Assumption \ref{asmp_community} holds and there exists at least one stubborn agent in the network (i.e., $n_{r}  < n$). The following results hold for the gossip model with stubborn agents.\\
(i) The model has a unique stationary distribution $\pi$ with mean $\mathbf{x}^r$, and $X^r(t) \overset{\textup{d}}{\to} \pi$, as $t \to \infty$. 
%Moreover, the convergence has the following rate
%\begin{align}\label{eq_wasserstein}
%     \limsup_{t\to \infty} \sup_{x \in \mathcal{S}} (d_W(P^t(x, \cdot), \pi))^{1/t} \le \rho(\bar{A}).
%\end{align}

(ii) The expectation of the state vector converges to $\mathbf{x}^r$:
\begin{align}\label{eq_expectation_limit}
    \mathbf{x}^r = \lim_{t \to \infty} \mathbb{E}\{X^r(t)\} = (I-\bar{A})^{-1}\bar{B}\mathbf{x}^s.
\end{align}
(iii) Denote $S^r(t) := \frac1t \sum_{i = 0}^{t - 1} X^r(i)$, then
\begin{align} \label{eq_ergodicity}
\lim_{t \to \infty} S^r(t) &= \mathbf{x}^r \quad \text{a.s.,} %\\%\label{eq_clt}
%sqrt{t} (S^r(t) - \mathbf{x}^r) &\overset{\textup{d}}{\to} \mathcal{N}(\mathbf{0},\mathbf{S}),\quad t \to \infty,
\end{align}
%where $\mathcal{N}(\mathbf{0},\mathbf{S})$ is the multivariate normal distribution with zero mean and covariance matrix
%\begin{align*}
%    \mathbf{S} = (I-\bar{A})^{-1} \mathbb{E}\left\{ 
%    Y \begin{bmatrix}
%    \mathbf{X} & \mathbf{x}^r\\
%    (\mathbf{x}^r)^T & 1
%    \end{bmatrix} Y^T
%    \right\} (I-\bar{A})^{-1},
%\end{align*}
%and
%\begin{align*}
%    Y &= [A(0) - \bar{A}~~ (B(0) - \bar{B})\mathbf{x}^s],\\
%    \mathbf{X} &= \lim_{t\to \infty} \mathbb{E} \{X^r(t)(X^r(t))^T\}.
%\end{align*}
\end{thm}

\begin{pf}
See Appendix \ref{append_proof_thm_stability}.\hfill$\Box$
\end{pf}

\begin{rem}
The first two results show that the agent states, although may not converge a.s., converge in distribution to a unique stationary distribution, and their expectations converge to the mean of the stationary distribution.
The third result indicates that we can obtain the value of $\mathbf{x}^r$ by computing the state time average.
\end{rem}

% In Theorem \ref{thm_stability}, $(I-\bar{A})^{-1}$ also has a block structure, as shown in the following proposition. This, combined with the above theorem, indicates that the states of regular agents in the same community is similar on average.

The next proposition shows that $\mathbf{x}^r$ also has a block structure, indicating that regular agents in the same community behave similarly on average.

\begin{prop}\label{thm_expectation_structure}
Under the conditions of Theorem~\ref{thm_stability}, 
\begin{align*}%\label{eq_inverse_I-barA}
    &(I - \bar{A})^{-1} = (1-q)^{-1}\\
    & \begin{bmatrix}
    \frac1{a_1}(I_{n_{r1}} + \tilde{w}_{s1} \mathbf{1}_{n_{r1},n_{r1}}) & \tilde{w}_d \mathbf{1}_{n_{r1},n_{r2}} \\
    \tilde{w}_d \mathbf{1}_{n_{r2},n_{r1}} & \frac1{a_2}(I_{n_{r2}} + \tilde{w}_{s2} \mathbf{1}_{n_{r2},n_{r2}})
    \end{bmatrix},
\end{align*}
where for $k=1,2$, $a_k = w_s n_k + w_d n_{3-k}$, $\tilde{w}_d = w_d / \delta$,
\begin{align}\nonumber
    \tilde{w}_{sk} &= (w_s^2 n_{s,3-k} + w_s w_d n_k + w_d^2 n_{r, 3-k}) / \delta,\\\nonumber
    \delta &= w_s^2n_{s1}n_{s2} + w_sw_d(n_1n_{s1}+n_2n_{s2}) \\\label{eq_defn_delta}
    &\quad + w_d^2(n_1n_2-n_{r1}n_{r2}).
\end{align}
As a result, 
\begin{align}\nonumber
    \mathbf{x}^r &= \frac{1}{\delta}[(\gamma_{11} \mathbf{1}_{n_{s1}}^T \mathbf{x}^{s1} + \gamma_{12} \mathbf{1}_{n_{s2}}^T \mathbf{x}^{s2}) \mathbf{1}_{n_{r1}}^T, \\\nonumber
    &\quad (\gamma_{21} \mathbf{1}_{n_{s1}}^T \mathbf{x}^{s1} + \gamma_{22} \mathbf{1}_{n_{s2}}^T \mathbf{x}^{s2}) \mathbf{1}_{n_{r2}}^T]^T\\\label{eq_chi1_chi2}
    &:=  [\chi_1 \mathbf{1}_{n_{r1}}^T, ~ \chi_2 \mathbf{1}_{n_{r2}}^T]^T,
\end{align}
where $\mathbf{x}^r$ is given in \eqref{eq_expectation_limit},
\begin{align*}
    \gamma_{kk} &= w_s^2n_{s,3-k} + w_sw_dn_k+ w_d^2n_{r, 3-k},\\
    \gamma_{k,3-k} &= w_d(w_sn_{3-k} + w_dn_k),%\\
    % \gamma_{11} &= w_s^2n_{s2} + w_sw_dn_1 + w_d^2n_{r2},\\
    % \gamma_{12} &= w_d(w_sn_2 + w_dn_1),\\
    % \gamma_{21} &= w_d(w_sn_1 + w_dn_2),\\
    % \gamma_{22} &= w_s^2n_{s1} + w_sw_dn_2 + w_d^2n_{r1},
\end{align*}
and $\mathbf{1}_{n_{sk}}^T \mathbf{x}^{sk}$ is defined to be zero if $n_{sk} = 0$, $k=1,2$.
\end{prop}

\begin{pf}
From \eqref{eq_barA}, we have that
\begin{align*}
    &I - \bar{A} = \\&(1-q) \begin{bmatrix}
    a_1 I_{n_{r1}} - w_s \mathbf{1}_{n_{r1},n_{r1}} & -w_d \mathbf{1}_{n_{r1},n_{r2}}\\
    -w_d \mathbf{1}_{n_{r2},n_{r1}} & a_2 I_{n_{r2}} - w_s \mathbf{1}_{n_{r2},n_{r2}}
    \end{bmatrix}.
\end{align*}
By Corollary \ref{cor_rho_barA}, $(I - \bar{A})^{-1}$ exists, and direct computation implies the first conclusion.
Hence, 
\begin{align*}
    (I - \bar{A})^{-1} \bar{B} = 
    \begin{bmatrix}
    \tilde{w}_{s1} \mathbf{1}_{n_{r1},n_{s1}} & a_2 \tilde{w}_d \mathbf{1}_{n_{r1},n_{s2}} \\
    a_1 \tilde{w}_d \mathbf{1}_{n_{r2},n_{s1}} & \tilde{w}_{s2} \mathbf{1}_{n_{r2},n_{s2}}
    \end{bmatrix}.
\end{align*}
Then under Assumption~\ref{asmp_community}, \eqref{eq_chi1_chi2} is obtained from~\eqref{eq_expectation_limit}.\hfill$\Box$
\end{pf}

\begin{rem}
%The explicit expressions of $\chi_1$ and $\chi_2$ are given in~\cite{xing2021detecting}. 
In Appendix~B, we study the block structure of $\mathbf{x}^r$ in a multiple-community case, as a generalization of Proposition~\ref{thm_expectation_structure}. 
\end{rem}

The above proposition means that 
%the limit of expectation of regular-agent states is a weighted average of stubborn agent states. %(Note that $\delta = \gamma_{11} n_{s1} + \gamma_{12} n_{s2} = \gamma_{21} n_{s1} + \gamma_{22} n_{s2}$). 
%Moreover, 
regular agents in the same community have the same limit, which is a weighted average of stubborn states.  Hence it is possible to split regular agents by computing the state time average. However, we are unable to do so if only one community has stubborn agents, or the stubborn states are similar. The following condition rules out these cases. 
%To partition regular agents according to their states, we introduce the following condition to ensure that $\chi_1$ and $\chi_2$ in Proposition \ref{thm_expectation_structure} are not equal. Otherwise, the regular agents exhibit similar behavior on average, revealing nothing about the community structure.

\begin{assum}\label{asmp_stubborn_agents} Both communities have stubborn~agents (i.e., $n_{s1} n_{s2} > 0$), and $\mathbf{x}^s = [(\mathbf{x}^{s1})^T ~ (\mathbf{x}^{s2})^T]^T$ satisfies that
$\mathbf{1}_{n_{s1}}^T \mathbf{x}^{s1}/n_{s1} \not = \mathbf{1}_{n_{s2}}^T \mathbf{x}^{s2}/n_{s2}$.
\end{assum}

This assumption has a practical meaning: stubborn agents are distributed among communities, and agents from different communities are more likely to have distinct opinions. Under Assumption~\ref{asmp_stubborn_agents}, we have the following result, indicating that the presence of stubborn agents enhances the separation of regular agents.

\begin{prop}\label{thm_identifiability}
Under the conditions of Theorem~\ref{thm_stability}, %for $\chi_1$ and $\chi_2$ given in \eqref{eq_chi1_chi2}, 
$\chi_1 \not= \chi_2$ if and only if Assumption~\ref{asmp_stubborn_agents} holds.
\end{prop}

\begin{pf}
It suffices to note from Proposition \ref{thm_expectation_structure}, 
\begin{align*}
    \chi_1 - \chi_2 &= \frac{1}{\delta}((\gamma_{11} - \gamma_{21}) \mathbf{1}_{n_{s1}}^T \mathbf{x}^{s1} + (\gamma_{12} - \gamma_{22}) \mathbf{1}_{n_{s2}}^T \mathbf{x}^{s2})\\
    &=\frac{1}{\delta} (w_s^2 - w_d^2) (n_{s2}\mathbf{1}_{n_{s1}}^T \mathbf{x}^{s1} - n_{s1}\mathbf{1}_{n_{s2}}^T \mathbf{x}^{s2}). \quad\hfill\Box
\end{align*}
\end{pf}

This result shows that Assumption~\ref{asmp_stubborn_agents} is a necessary and sufficient condition for regular agents from different communities having nonidentical expected stationary states. Note that $\mathbf{1}_{n_{s1}}^T \mathbf{x}^{s1}/n_{s1} \not = \mathbf{1}_{n_{s2}}^T \mathbf{x}^{s2}/n_{s2}$ is generic (i.e., it holds for almost all $\mathbf{x}^s \in \mathbb{R}^{n_s}$).

%\begin{rem}
%The above result illustrates an intuitive but crucial fact that the averages of stubborn agent states in the two communities must not be identical. Otherwise, their influence on regular agents would be the same.
%\end{rem}

\section{Joint {\color{black}Recovery} and Estimation Algorithm}\label{sec_alg}
In this section, we design a joint {\color{black}recovery} and estimation algorithm (Algorithm~\ref{alg_1}) to address the considered problem. 
We assume the following connections between stubborn and regular agents are known. The information means that we have prior knowledge about stubborn agents, which may be gathered from other sources in practice.  

%Since there is no information for the community labels of stubborn agents which do not update, we assume the following connections between stubborn and regular agents are known for Algorithm~\ref{alg_1}. Intuitively, it means that we have some prior knowledge of stubborn agents' labels.

\begin{assum}\label{asmp_stubborn_agent_structure}
For every stubborn agent $i \in \mathcal{V}_s$, it is known for Algorithm~\ref{alg_1} that there exists a regular agent $j_i \in \mathcal{V}_r$ such that $i$ and $j_i$ are in the same community (i.e, $\mathcal{C}(i) = \mathcal{C}(j_i)$).
\end{assum}

Now we are ready to introduce Algorithm~\ref{alg_1}, in which we denote the estimates at time $t$ of community label $\mathcal{C}(i)$, interaction probabilities $w_s$ and $w_d$, by $\hat{\mathcal{C}}(i,t)$, $\hat{w}_s(t)$, and $\hat{w}_d(t)$, respectively. %, $i \in \mathcal{V}$. 
In addition, %since we have sorted agents as in Section~\ref{sec_model}, for a regular agent~$i \in \mathcal{V}_2$, the $(i - n_1 + n_{r1})$-th component of $S^r(t)$ should be used to generate the estimates. But we write it as $S^r_i(t)$ in the algorithm, with a slight abuse of notation, to emphasize that $n_1$ and $n_{r1}$ are unknown in the task.
we use $S^r_i(t)$ to represent the $(i-n_1+n_{r1})$-th entry of $S^r(t)$, $i \in \mathcal{V}_2 = \{n_1+1, \dots, n_1+n_{r2}\}$ for simplicity. Note that both $n_1$ and $n_{r1}$ are unknown in the algorithm. {\color{black}In the gossip model, agents randomly interact and update states. Algorithm~\ref{alg_1} partitions the agents and estimates interaction strength between agents, out of these state observations, without interaction information.}

\begin{algorithm}[t]
\caption{(Joint {\color{black}Recovery} and Estimation)}
\label{alg_1}
\small \textbf{Input:} $\{X^r(t), t=0,1,2,\dots\}$,  $X^s(0)$, step-size parameter~$a$ of the interaction estimator with $a > 0$.\\
\textbf{Output:} $\{\hat{\mathcal{C}}(i,t)\}$, $\hat{w}_s(t)$, $\hat{w}_d(t)$.
\begin{algorithmic}[1]
\STATE{Randomize $\hat{\mathcal{C}}(i,0)$, $\hat{w}_s(0)$, $\hat{w}_d(0)$, set $S^r(0) = X^r(0)$. }
\FOR{$t = 1, \dots$}
\STATE{Compute
\begin{align*}
    S^r(t) &= \frac{t}{t+1} S^r(t-1) + \frac{1}{t+1} X^r(t),\\
    \bar{s}^r(t) &= \frac{1}{n_r}\mathbf{1}_{n_r}^T S^r(t).
\end{align*}
}
\STATE{\textbf{Community recovery:}
\begin{align*}
    \hat{\mathcal{C}}(i, t) &= 2 - \mathbb{I}_{[S^r_i(t) > \bar{s}^r(t)]}, ~i \in \mathcal{V}_r,\\
    \hat{\mathcal{C}}(i, t) &= \hat{\mathcal{C}}(j_i, t), ~i \in \mathcal{V}_s,
\end{align*}
where $j_i$ is defined in Assumption \ref{asmp_stubborn_agent_structure}.
}
\STATE{\textbf{Interaction estimation:}
\begin{align*}
    \hat{w}_s(t) &= \hat{w}_s(t-1) -  \frac{a}{t} \text{sgn}(g(t)) \Big(g(t) \hat{w}_s(t-1) \\
    &\quad+ \frac{h_2(t)}{\hat{n}_1(t)\hat{n}_2(t)}\Big),\\
    % \hat{w}_s(t) &= \tilde{w}_s(t) \mathbb{I}_{[|\tilde{w}_s(t)| < 2]} + \frac12 \mathbb{I}_{[|\tilde{w}_s(t)| \ge 2]},\\
    \hat{w}_d(t) &= \frac{2 - \hat{w}_s(t) (\hat{n}_1^2(t) + \hat{n}_2^2(t) - \hat{n}_1(t) - \hat{n}_2(t))}{2\hat{n}_1(t)\hat{n}_2(t)},
\end{align*}
where 
\begin{align*}
    &g(t) = h_1(t) - \frac{\hat{n}_1^2(t) + \hat{n}_2^2(t) - \hat{n}_1(t) - \hat{n}_2(t)}{2\hat{n}_1(t)\hat{n}_2(t)} h_2(t),\\
    &h_1(t) = \frac{|\hat{\mathcal{V}}_{s1}(t)|}{|\hat{\mathcal{V}}_{r1}(t)|} \textstyle\sum_{i\in \hat{\mathcal{V}}_{r1}(t)} S^r_i(t) - \textstyle\sum_{i\in \hat{\mathcal{V}}_{s1}(t)} X^s_i(0),\\
    &h_2(t) = \frac{\hat{n}_2(t)}{|\hat{\mathcal{V}}_{r1}(t)|} \textstyle\sum_{i\in \hat{\mathcal{V}}_{r1}(t)} S^r_i(t)\\
    &\qquad\qquad  - \textstyle\sum_{i\in \hat{\mathcal{V}}_{r2}(t)} S^r_i(t) - \textstyle\sum_{i\in \hat{\mathcal{V}}_{s2}(t)} X^s_i(0),\\
    &\hat{n}_k(t) = \textstyle\sum_{i \in \mathcal{V}} \mathbb{I}_{[\hat{\mathcal{C}}(i,t) = k]},\\
    &\hat{\mathcal{V}}_{rk}(t) = \{i \in \mathcal{V}_r : \hat{\mathcal{C}}(i,t) = k\},\\
    &\hat{\mathcal{V}}_{sk}(t) = \{i \in \mathcal{V}_s : \hat{\mathcal{C}}(i,t) = k\}, ~k=1,2.
\end{align*}
}
\ENDFOR
\end{algorithmic}
\end{algorithm}

\begin{rem}
{\color{black}The difficulty of recovery is to find a quantity revealing the community structure. Algorithm~\ref{alg_1} exploits the trajectory data by using $S^r(t)$. From Proposition~\ref{thm_expectation_structure} we know that the entries of $S^r(t)$ converge to two distinct values corresponding to the communities. Hence clustering methods (Line~4 of Algorithm~\ref{alg_1}, or other methods such as $k$-means) can be used. For estimation of interaction probabilities, the key is to find consistent parameter equations. Here we use the property of stationary states. Note that, from \eqref{eq_expectation_limit}, it follows that $\mathbf{x}^r$ satisfies the following equation, 
\[
    \mathbf{x}^r = \bar{A} \mathbf{x}^r + \bar{B} \mathbf{x}^s,
\]
which implies that
\[
    w_s(n_{s1}\chi_1 - \mathbf{1}_{n_{s1}}^T \mathbf{x}^{s1}) + w_d(n_{2}\chi_1 - n_{r2}\chi_2 - \mathbf{1}_{n_{s2}}^T \mathbf{x}^{s2}) = 0.
\]
Assumption~\ref{asmp_stubborn_agents} ensures $n_{s1}\chi_1 - \mathbf{1}_{n_{s1}}^T\mathbf{x}^{s1} \not= 0$. Hence, $(n_{s1}\chi_1 - \mathbf{1}_{n_{s1}}^T \mathbf{x}^{s1})(n_{2}\chi_1 - n_{r2}\chi_2 - \mathbf{1}_{n_{s2}}^T \mathbf{x}^{s2}) < 0$ from $w_s w_d > 0$, under Assumptions~\ref{asmp_community} and~\ref{asmp_stubborn_agents}. From the definition of $W$, $w_s$ and $w_d$ also satisfy the relation given in Assumption~\ref{asmp_community}~(ii). Therefore, the following system of linear equations for~$(x~y)^T$
\begin{align}\label{eq_linear_system}
    \begin{cases}
    (n_{s1}\chi_1 - \mathbf{1}_{n_{s1}}^T \mathbf{x}^{s1})x + (n_{2}\chi_1 - n_{r2}\chi_2 - \mathbf{1}_{n_{s2}}^T \mathbf{x}^{s2})y = 0\\
    (n_1(n_1 - 1) + n_2(n_2 - 1)) x + 2 n_1 n_2 y = 2
    \end{cases}
\end{align}
has a unique solution $(w_s ~ w_d)$, under Assumptions~\ref{asmp_community} and~\ref{asmp_stubborn_agents}, for fixed $n_k$, $n_{rk}$, and $n_{sk}$. But these quantities are unknown, so we leverage SA techniques to estimate them, as presented in Line~5 of Algorithm~\ref{alg_1}. Note that the algorithm does not need to know the averaging weight $q$.}
%It computes the time average, and then uses it to cluster the agents into two groups. 
%%The idea is to compare the entries of this time average. Other clustering methods may be used to solve this problem, but the clustering result must have theoretical guarantees, to ensure the estimates of $w_s$ and $w_d$ are convergent. Otherwise, incorrect knowledge of community structure could result in inconsistent estimation. 
%In the algorithm, we use a simple rule for community {\color{black}recovery}, to ensure convergence of Algorithm~\ref{alg_1}. In practice, other clustering methods such as $k$-means may be applied. We will study their performance and their influence on the performance of the interaction estimator in the future.
\end{rem}

\section{Convergence Analysis}\label{sec_alg_analysis}
This section studies the performance of Algorithm~\ref{alg_1}.
The following result means that communities can be recovered in finite time, and the interaction probability estimates are convergent.

\begin{thm}\label{thm_convergence}(Convergence of Algorithm~\ref{alg_1})~\\
Under Assumptions \ref{asmp_community}-\ref{asmp_stubborn_agent_structure}, the following holds.\\
(i) The community {\color{black}recovery} is achieved in finite time: there exists a positive integer-valued random variable $T$ such that $\hat{\mathcal{C}}(i,t) = \mathcal{C}(i)$, for all $i \in \mathcal{V}$ and $t > T$.\\
(ii) The interaction estimator converges a.s., namely,
\begin{equation*}
    \mathbb{P}\left\{\lim_{t \to \infty}(\hat{w}_s(t), \hat{w}_d(t)) = (w_s, w_d)\right\} = 1.
\end{equation*}
% for any constant $a > 0$ given in Line~5 of Algorithm~\ref{alg_1}, where $\hat{w}_s(t)$ and $\hat{w}_d(t)$ are the estimates of $w_s$ and $w_d$ at time $t$.
\end{thm}

\begin{pf}
See Appendix \ref{append_proof_thm_convergence}.\hfill$\Box$
\end{pf}

\begin{rem}\label{rem_ergodic}
Since Algorithm~\ref{alg_1} uses the property~\eqref{eq_ergodicity}, it can also deal with situations where state observations are corrupted. For example, one cannot observe the whole trajectory but can only sample the states at some time steps. Ergodic property ensures that the time average of the sampled states still converges, if the sampling process is independent of the update, and the number of samples tends to infinity \cite{wai2016active, ravazzi2014ergodic}. Another situation is that the observations are disturbed by i.i.d. zero-mean noise independent of the process. The law of large numbers guarantees that the influence of noise vanishes over time. %Performance of Algorithm~\ref{alg_1} under these situations is illustrated via simulation in~\cite{xing2021detecting}.
\end{rem}

Now we investigate the sample complexity of the community recovery, and the convergence rate of the interaction estimator. The following result is useful for studying the sample complexity of the community {\color{black}recovery}.

\begin{lem}\label{lem_concentration}
Consider a Markov chain $\{X(t)\}$ taking values on a compact state space $\mathcal{X}$ and having a unique stationary distribution $\pi$. For a function $f: \mathcal{X} \to \mathbb{R}$ and $\alpha := \int_{\mathcal{X}} f(x) \pi(dx)$, denote $g(x) := \sum_{t=0}^{\infty} \mathbb{E}\{f(X(t)) - \alpha | X(0) = x\}$, and the supremum of $|g|$ on $\mathcal{X}$ by $\|g\|_s := \sup\{|g(x)|:x\in \mathcal{X}\}$.
%\begin{align*}
%    g(x) &:= \sum_{t=0}^{\infty} \mathbb{E}\{f(X(t)) - \alpha | X(0) = x\},\\
%    \|g\|_s &:= \sup\{|g(x)|:x\in \mathcal{X}\}.
%\end{align*}
If $\|g\|_s < \infty$, then, for all $\varepsilon > 0$ and $t > 2 \|g\|_s/\varepsilon$, it holds for $S_f(t) := \frac1t \sum_{i=0}^{t-1}f(X(i))$ that
\begin{equation*}
    \mathbb{P}\{|S_f(t) - \alpha| \ge \varepsilon\} \le 2 \exp\left\{-\frac{(t\varepsilon - 2 \|g\|_s)^2}{2 t \|g\|_s^2}\right\}.
\end{equation*}
\end{lem}

\begin{pf}
	See Appendix~\ref{append_proof_lem_concentration}.
\end{pf}

\begin{rem}
	{\color{black}Similar concentration results to~Lemma~\ref{lem_concentration} have been obtained in the literature for other models. One class of results leverage Markov chain approaches and normally require stability such as uniform ergodicity~\cite{glynn2002hoeffding,paulin2015concentration} or explicit bounds of the derivative of the initial measure with respect to the stationary measure~\cite{fan2021hoeffding}. It is hard to derive these properties for Markov chains without continuous distributions~\cite{gibbs2002choosing}, as in our case. Another line of research studies concentration of Polyak averages, and contains step-size conditions~\cite{mou2020linear}, which cannot be applied to our problem either.}
\end{rem}

{\color{black}Using the preceding lemma, we are able to compute when the differences between entries of $S^r(t)$ and $\mathbf{x}^r$ are small enough, such that agents in different communities have distinct state time averages. As a result, we obtain a sample-complexity result for the community recovery.} The next theorem shows that the probability of {\color{black}recovering} communities successfully depends on the network, the interaction probabilities, and the stubborn-agent states. The probability tends to one as~$t$ goes to infinity. 

\begin{thm}(Sample complexity)\label{thm_finitesample}~\\
Under the conditions of Theorem~\ref{thm_convergence}, for the community {\color{black}recovery} step of Algorithm~\ref{alg_1}, it holds that, for $t > t_0$,
\begin{equation*}
    \mathbb{P}\big\{ \hat{\mathcal{C}}(i,t) = \mathcal{C}(i), \forall i \in \mathcal{V} \big\} \ge 1 - 2n_r \exp\bigg\{\frac{-2(t - t_0)^2}{t_0^2 t}\bigg\},
\end{equation*}
with %\[t_0 = \frac{4 \delta (2-\rho(\bar{A})) n_r^{\frac32} (n_r+1)  \max\{|\underline{s}|,|\bar{s}|\}}{(1 - \rho(\bar{A}))^2~|w_s^2-w_d^2|~|n_{s1} \mathbf{1}^T_{n_{s2}} \mathbf{x}^{s2} - n_{s2} \mathbf{1}^T_{n_{s1}} \mathbf{x}^{s1}|}\]
$t_0 = 4 \delta c_{\bar{A}} c_{n_r} c_s/ c_w$, where $c_{\bar{A}} =1/$$(1 - \rho(\bar{A}))$, 
$c_{n_r} = n_r^{3/2} (n_r+1)$, 
$c_s = \max\{|\underline{s}|,|\bar{s}|\}/|n_{s1} \mathbf{1}^T_{n_{s2}} \mathbf{x}^{s2} - n_{s2} \mathbf{1}^T_{n_{s1}} \mathbf{x}^{s1}|$,
$c_w = |w_s^2-w_d^2|$, $\delta$ is given in Proposition~\ref{thm_expectation_structure}, and $\underline{s}$ and $\bar{s}$ are given in~\eqref{eq_mathcalS}.
\end{thm}

\begin{figure*}
    \centering
%    \subfigure[\label{fig_xrt}State evolution of four selected regular agents in the network.]{\includegraphics[width=0.3\linewidth]{xrt.pdf}}\qquad\qquad
%    \subfigure[\label{fig_srt}Convergence of the time average of states. Each line corresponds to one agent, and the four thick lines correspond to the selected agents in~(a).]{\includegraphics[width=0.3\linewidth]{srt.pdf}}\\
    \subfigure[\label{fig_community_detection}Finite-time community recovery. The recovery is achieved after $t=383$.]{\qquad\qquad
    \includegraphics[width=0.25\linewidth]{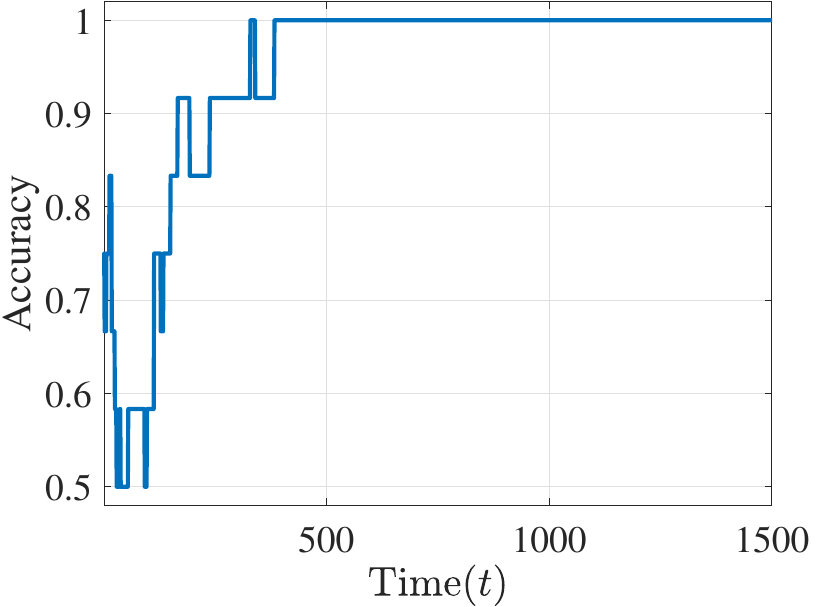}\qquad\qquad}\qquad
    \subfigure[\label{fig_parameter_estimation}Convergence of the interaction estimator. The solid (dashed) lines are true values (estimates).]{\qquad\qquad
    \includegraphics[width=0.25\linewidth]{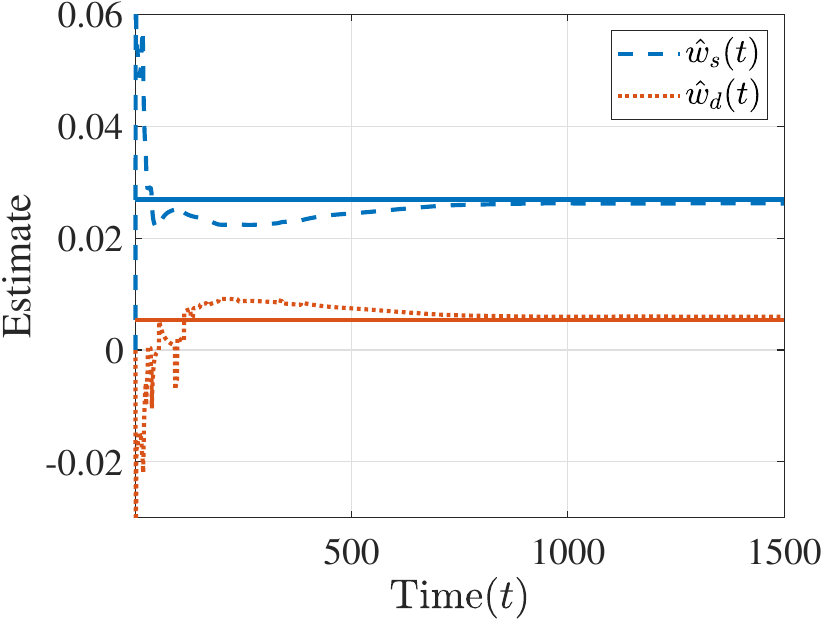}\qquad\qquad}
    \caption{\label{fig_1}Performance of Algorithm \ref{alg_1}.}
\end{figure*}

\begin{pf}
See Appendix \ref{append_proof_thm_finitesample}.\hfill$\Box$
\end{pf}

\begin{rem}\label{rem_thm_finitesample}
This result provides a sample complexity characterization for {\color{black}recovering} community from a single trajectory. Multiple-trajectory sample complexity is investigated by~\cite{wai2019blind,schaub2020blind,roddenberry2020exact}. The parameter $\delta$ reflects the combined effect of the cardinality of stubborn and regular agents and the interaction probabilities. $c_{\bar{A}}$ captures the ``speed'' of information diffusion, and increases with $\rho(\bar{A})$. $c_{n_r}$ depends on the number of regular agents. $c_s$ increases with the range of the states and decreases with the difference of averaged stubborn states in different communities, and $c_w$ measures the difference between interaction probabilities within and between communities. Smaller $\delta$, $c_{\bar{A}}$, $n_r$, and $\max\{|\underline{s}|,|\bar{s}|\}$ would make the {\color{black}recovery} easier, and so would larger $c_w$ and $|n_{s1} \mathbf{1}^T_{n_{s2}} \mathbf{x}^{s2} - n_{s2} \mathbf{1}^T_{n_{s1}} \mathbf{x}^{s1}|$. {\color{black}For the gossip model over a graph sampled from an SBM, Example~\ref{exmp_sbm} indicates that the algorithm can recover most of the community labels, which is illustrated in Section~\ref{sec_simulation}.}
\end{rem}

We have the following result for the convergence rate of the interaction estimator. It shows that the convergence rate also depends on the model parameters, and a large enough step-size parameter $a$ ensures that the rate can achieve $O(1/\sqrt{t})$.

\begin{thm}(Convergence rate)\label{thm_convergencerate}~\\
Under the conditions of Theorem~\ref{thm_convergence}, %for the interaction estimator of Algorithm~\ref{alg_1}, 
it holds for $d_0 \in [0, \min\{1/2,a|\eta|\})$ that
\begin{align*}
    (\hat{w}_s(t) - w_s, \hat{w}_d(t) - w_d) = o(t^{-d_0}), \text{ a.s.}, %\\ \forall ),
\end{align*}
where $a>0$ is the step-size parameter given in Algorithm~\ref{alg_1}, $\eta = (w_sn_2 + w_dn_1) (n_{s1} \mathbf{1}^T_{n_{s2}} \mathbf{x}^{s2} - n_{s2} \mathbf{1}^T_{n_{s1}} \mathbf{x}^{s1})/$ $(\delta n_1n_2)$, and $\delta$ is given in Proposition~\ref{thm_expectation_structure}. 
\end{thm}

\begin{pf}
See Appendix \ref{append_proof_thm_convergencerate}.\hfill$\Box$
\end{pf}

\begin{rem}
In the theorem, $\eta$ increases with the combined effect of the number of agents and the interaction probabilities (i.e., $(w_sn_2+w_dn_1)/\delta$) and with the disagreement of stubborn agents, and decreases with the cardinality of each community. When $a \ge 1/(2|\eta|)$, the estimator achieves its optimal rate. Larger $\eta$ provides a wider selection range. {\color{black}Simulation in Section~\ref{sec_simulation} shows that the algorithm using a trajectory from the gossip model over an SBM can estimate the ratio of the link probabilities.}
\end{rem}

%%%%%%%%%%%%%%%%%%%%%%%%%%%%%%%%%%%%%%%%%%%%%%%%%%%%%%%%%%%%%%%%%%%%%%%%%%%%%%%%

\section{Numerical Simulation}\label{sec_simulation}
This section illustrates the performance of Algorithm~\ref{alg_1}, conducts an algorithm comparison, and applies Algorithm~\ref{alg_1} to the SBM case and a real network.

%\subsection{Performance of Algorithm \ref{alg_1}}\label{sec_simul_converg}
To illustrate the performance of Algorithm~\ref{alg_1} under Assumptions \ref{asmp_community}-\ref{asmp_stubborn_agent_structure}, consider a network consisting of twelve agents. The two communities both have five regular agents and one stubborn agent. Set interaction probabilities be $w_s = 5/186$ and $w_d =~1/186$.
The stubborn agent in community~$1$ (resp. community~$2$) has state $1$ (resp. $-1$). The initial states of regular agents are drawn from uniform distribution on $(-1,1)$. The averaging weight is set to be $q = 1/2$ in all experiments.
%Fig.~\ref{fig_xrt}-(b) demonstrate the state evolution of four regular agents (half of them are in community~$1$ and the other half are in community~$2$), and the time average of regular agent states. The subfigures show that it is not easy to determine the community structure by directly observing the states, but the time average of states reveals the underlying communities. 
Fig.~\ref{fig_community_detection} shows that Algorithm~\ref{alg_1} recovers the communities in finite time, where the accuracy at time~$t$ is defined by $\frac1n (\max_{\sigma \in S_2}  \{  \sum_{i=1}^n \mathbb{I}_{[\sigma(\hat{\mathcal{C}}(i,t)) = \mathcal{C}(i)]} \}) \in [0,1]$. Here $\sigma : \{1, 2\} \to \{1, 2\}$ is a permutation function (to prevent a reverse distribution of labels), $S_2$ is the group of permutations on $\{1, 2\}$, $\mathcal{C}(i)$ is agent $i$'s community label, $\hat{\mathcal{C}}(i,t)$ is the estimate of agent $i$'s label at time $t$, and $n = 12$. Consistency of the interaction estimator with step-size parameter $a = 1$ is demonstrated in Fig.~\ref{fig_parameter_estimation}. 
These results validate Theorem~\ref{thm_convergence}.

We now show the sample complexity of the community {\color{black}recovery} (Theorem~\ref{thm_finitesample}) and compare the recovery step with the $k$-means, $k$-means++ \cite{arthur2006k}, and spectral clustering methods \cite{%von2007tutorial,
abbe2017community}. This experiment considers the gossip model under Assumptions \ref{asmp_community}-\ref{asmp_stubborn_agent_structure} with $n = 400$, $n_1 = 150$, and $n_{s1} = n_{s2} = 8$. Let $w_s/w_d = 5$ and solve the two parameters from~\eqref{eq_ws_wd}. Let stubborn agents in community~$1$ (resp. community~$2$) have state $1$ (resp. $-1$), and generate the initial states of other agents from uniform distribution on $(-1,1)$.
By running the algorithms for $200$ times, we obtain the relative frequency that the algorithms {\color{black}recover} all community labels, defined by $p_t := (\sum\nolimits_{k=1}^N \max_{\sigma \in S_2} \{ \mathbb{I}_{\left[\sigma(\hat{\mathcal{C}}_k(i,t)) = \mathcal{C}(i), \forall i \in \mathcal{V} \right]} \})/N$, where $N=200$ and $\hat{\mathcal{C}}_k(i,t)$ is the estimate of agent~$i$'s label at time $t$ in the $k$-th run.
%Simulation is conducted by using $200$ runs of the same model as in Section~\ref{sec_simul_converg}. 
After computing the time average $S^r(t)$, we use $k$-means and $k$-means++ with $k=2$ instead of Line~4 of Algorithm~\ref{alg_1}, to recover communities. To implement the spectral clustering method,  assume that edge activation is known, and use the activation information to estimate the interaction probability matrix $W$. Applying spectral clustering to estimates of $W$ obtains community estimates. 

\begin{figure}
    \centering
%    \subfigure[The accuracy of community detection by three algorithms in one run.]{
%    \includegraphics[width=0.25\linewidth]{alg_comp.pdf}}\qquad\qquad
%    \subfigure[The relative frequency of incorrect community detection.]
	\includegraphics[width=0.68\linewidth]{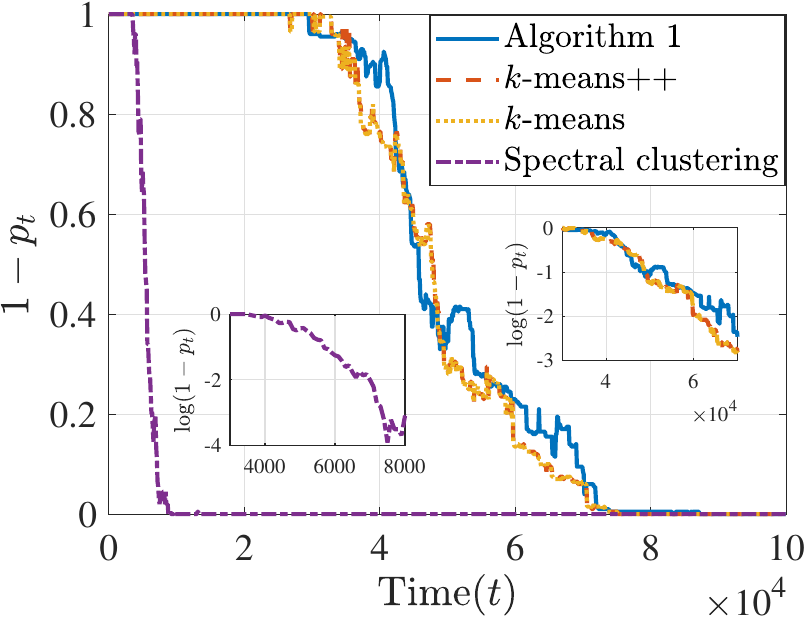}
    \caption{\label{fig_comp}Performance comparison of four methods.}
\end{figure}

Fig.~\ref{fig_comp} shows that the probabilities of unsuccessful community recovery of all approaches tends to zero exponentially over time. The spectral clustering method performs much better than other algorithms, because it directly uses interaction information, but the required time is still of the same order as the other algorithms. The $k$-means and $k$-means++ methods perform similarly to each other, and also similarly to Algorithm~\ref{alg_1}. This observation indicates that the major challenge of the considered problem is how to use agent states to recover communities without topological information.

\begin{figure*}
    \centering
    \subfigure[\label{fig_sbm_acc}Averaged accuracy of community recovery for each SBM. ]{\qquad\qquad\qquad
    \includegraphics[width=0.25\linewidth]{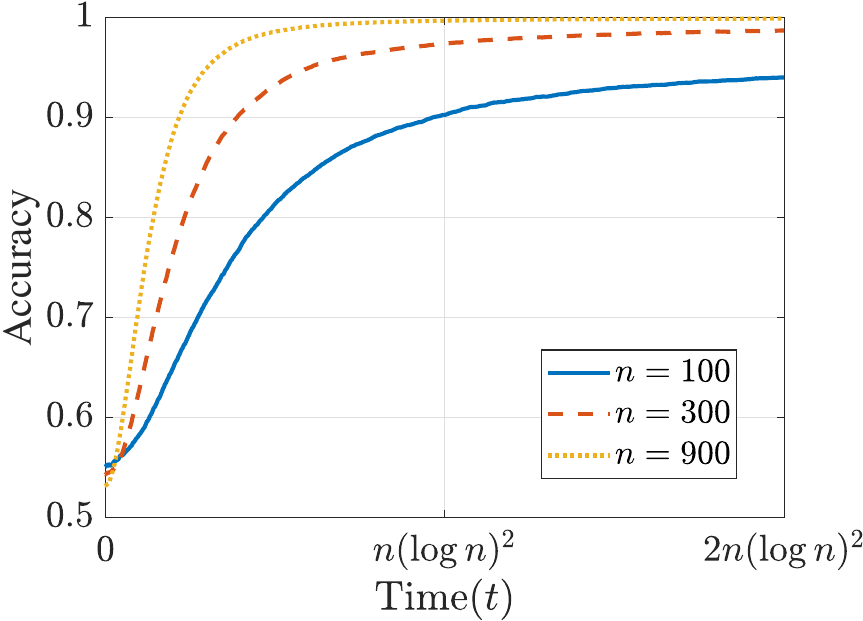}\qquad\qquad\qquad
    }~
    \subfigure[\label{fig_sbm_err} Median of estimation error of $p_s/p_d$ for each SBM.]{\qquad\qquad\qquad
    \includegraphics[width=0.25\linewidth]{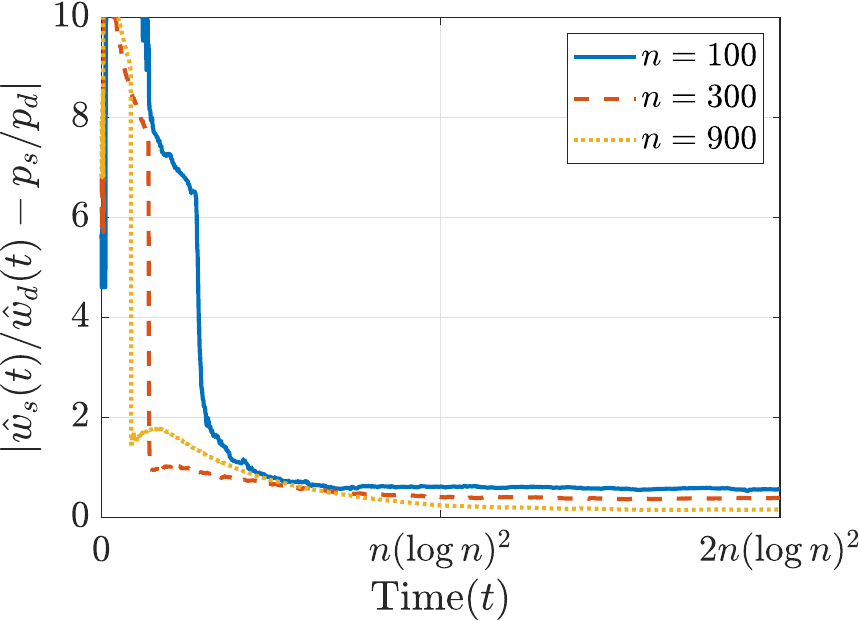}\qquad\qquad\qquad
    }
%    \subfigure[\label{fig_prediction}The evolution of $\bar{s}^r_k(t)$, $k=1,2$, during two conflicts. The dashed blue and red lines illustrate prediction of final positions of the two communities in the new conflict.]{
%    \includegraphics[width=0.5\linewidth]{prediction.pdf}
%    }
    \caption{\label{fig_sbm}Performance of Algorithm~\ref{alg_1} using trajectories of the gossip model over sampled graphs from SBMs with $n=100,300, 900$.}
\end{figure*}

\begin{figure*}
    \centering
    \subfigure[\label{fig_zachary_network}The community structure of Zachary's karate club network. Red squares and green triangles show two communities.]{\qquad\qquad\qquad
    \includegraphics[width=0.25\linewidth]{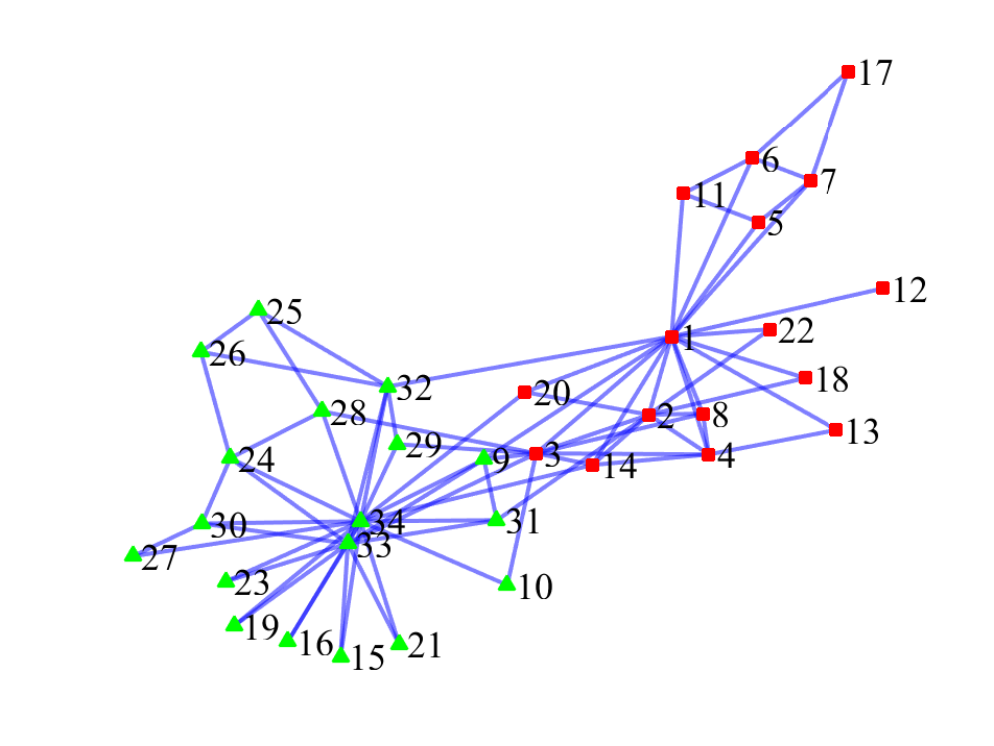}\qquad\qquad\qquad
    }~
    \subfigure[\label{fig_accuracy}Accuracy of community recovery of Algorithm~\ref{alg_1} for the gossip model over Zachary's karate club network.]{\qquad\qquad\qquad
    \includegraphics[width=0.25\linewidth]{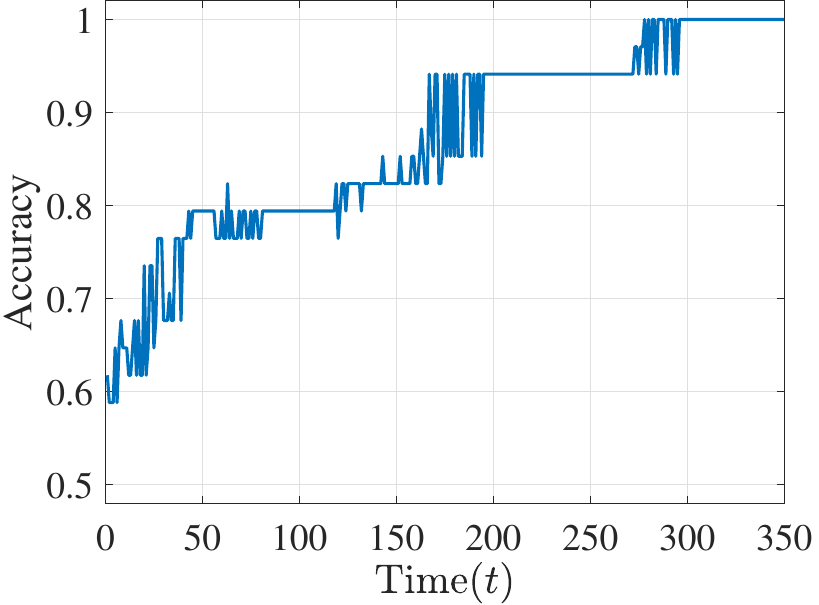}\qquad\qquad\qquad
    }
%    \subfigure[\label{fig_prediction}The evolution of $\bar{s}^r_k(t)$, $k=1,2$, during two conflicts. The dashed black and red lines illustrate prediction of final positions of the two communities in the new conflict.]{
%    \includegraphics[width=0.5\linewidth]{prediction.pdf}
%    }
    \caption{\label{fig_2}Numerical experiment over Zachary's karate club network.}
\end{figure*}

%\subsection{Numerical experiment over Zachary's karate club network}\label{sec_simul_real}

{\color{black}We now consider the case where trajectories of the gossip model over graphs sampled from SBMs are given to Algorithm~\ref{alg_1}. We use three SBMs with size $n=100, 300, 900$ and with two equal-sized communities ($\nu_1=\nu_2=0.5$). Set $n_{r1} = n_{r2} = 0.45n$, and $n_{s1} = n_{s2} = 0.05n$. Let the link probability in the same community be $p_s = (\log n)^2/n$ and the link probability between different communities be $p_d = (\log n)/n$. For each SBM, we generate $20$ graph samples. For each graph sample, we run Algorithm~\ref{alg_1} for $20$ times. Regular states are generated the same as earlier and stubborn agents in community~$1$ (resp. community~$2$) have state $1$ (resp. $-1$). Fig.~\ref{fig_sbm_acc} shows that Algorithm~\ref{alg_1} has high community recovery accuracy, increasing with $n$. This phenomenon results from the concentration discussed in Example~\ref{exmp_sbm}.  Algorithm~\ref{alg_1} outputs $\hat{w}_s(t)$ and $\hat{w}_d(t)$ as estimates of the two distinct non-zero values of $\mathbb{E}\{\mathcal{W}\}/\mathbb{E}\{\alpha\}$. Note that $[cp_s, cp_d]$ defines the same $\mathbb{E}\{\mathcal{W}\}/\mathbb{E}\{\alpha\}$ for all $c > 0$, so we can only estimate the ratio $p_s/p_d$ without knowing the expected number of edges of the SBM. Fig.~\ref{fig_sbm_err} shows that the median of the estimation error for trajectory samples from each SBM is close to zero and decreases with $n$.}

%\subsection{Numerical experiment over Zachary's karate club network}\label{sec_simul_real}

Zachary's karate club network \cite{zachary1977information}, presented in Fig. \ref{fig_zachary_network}, is used to demonstrate an application of Algorithm~\ref{alg_1}. An edge represents frequent interaction between the two agents. The strength of interactions between agents is modeled by a weighted adjacency matrix (see matrix $C$ in \cite{zachary1977information}). A conflict between agents $1$ and $34$ results in a fission of the club.
In the experiment, we assume that only the opinions can be observed, instead of interactions between agents. The process is modeled by the gossip model with stubborn agents. Agents $1$ and~$34$ are set to be stubborn agents holding different opinions. In addition, one edge in Fig.~\ref{fig_zachary_network} is selected at each time with a probability proportional to interaction strength given in \cite{zachary1977information}. %, and the two agents corresponding to this edge interact and exchange opinions. 
The goal is to partition the agents into communities based on only state observations. Note that the network structure departures from our assumptions, but the result shown in Fig.~\ref{fig_accuracy} indicates that our
algorithm can finally recover the community structure as time increases, without topological and interaction information.

\section{Conclusion and Future Work}\label{sec_conclusion}
In this paper, we {\color{black}developed a joint algorithm to recover the community structure and to estimate the interaction probabilities for gossip opinion dynamics}. It was proved that the community {\color{black}recovery} is achieved in finite time, and the interaction estimator converges almost surely. We analyzed the sample complexity of the recovery and convergence rate of the estimator. Future work includes  to study the case where all regular agents have the same stationary expectation, and to analyze {\color{black}the community detection problem for dynamics over the SBM}.

\begin{ack}                               % Place acknowledgements
This work was supported by Knut \& Alice Wallenberg Foundation, Swedish Research Council, and National Natural Science Foundation of China (Grant No. 11931018). %The authors thank the anonymous reviewers for their insightful comments and suggestions.  % here.
\end{ack}

\newpage

\appendix

\textbf{Appendix}

\section{Proof of Example~1}\label{append_proof_exmp1}
{\color{black}In this section, we prove the inequality given in Example~1. We need the following two results, whose proofs can be found in Sections~2.3 and~4.5 of~\cite{vershynin2018high}, respectively.

\begin{lem}[Chernoff inequality]\label{lem_chernoff_pf_thm_state_concentration}
    Let $X = \sum_{i=1}^n X_i$, where $X_i$ are independent Bernoulli random variables with expectation $p_i$, and let $\mu := \EE\{X\} = \sum_{i=1}^n p_i$. Then for all $\delta \in (0,1)$,
    \begin{align*}
        \PP\{|X-\mu| \ge \delta \mu\} \le 2 e^{-\mu \delta^2/3},
    \end{align*}
    and for all $a > \mu$, 
    \begin{align*}
        \PP\{X \ge a\} \le e^{-\mu} \Big(\frac{e \mu}{a}\Big)^a.
    \end{align*}
\end{lem}

\begin{lem}[Matrix Bernstein inequality]\label{lem_bernstein_pf_thm_state_concentration}
    Let $Y_i$, $1\le i\le k$, be independent $n$-dimensional random symmetric matrices such that $\|Y_i - \EE\{Y_i\}\| \le C_0$. Denote $Y := \sum_{i=1}^k Y_i$. Then for all $a > 0$, 
    \begin{align*}
        \PP\{\|Y - \EE\{Y\}\| > a\} \le 2n \exp\bigg\{\frac{-a^2}{2v^2 + 2C_0 a/3}\bigg\},
    \end{align*}
    where $v^2 = \|\sum_{i=1}^k \textup{var}(Y_i)\|$.
\end{lem}

Now we prove the inequality given in Example~1. For two sequences of real numbers $\{a_k\}$ and $\{b_k\}$ with $b_k \not= 0$, $k \ge 1$, denote $a_k \sim b_k$,  if $\lim_{k\to \infty} a_k/b_k = 1$. Hence for $\alpha =  \sum_{i=1}^n \sum_{j=i+1}^n a_{ij}$, it holds that
\begin{align*}
	\mathbb{E}\{\alpha\} &= \Big[\frac12 p_s \Big(\nu_1^2 +\nu_2^2 - \frac{\nu_1}{n} - \frac{\nu_2}{n}\Big) + p_d \nu_1 \nu_2\Big] n^2\\
	&\sim \Big[\frac12 p_s \Big(\nu_1^2 +\nu_2^2 \Big) + p_d \nu_1 \nu_2\Big] n^2.
\end{align*}
Note that
\begin{align*}
    &\bigg \|\tilde{W} - \frac{\mathbb{E}\{\mathcal{A}\}}{\mathbb{E}\{\alpha\}}  \bigg\| \\
    &= \bigg\| \frac{\mathcal{A}}{\alpha}  - \frac{\mathbb{E}\{\mathcal{A}\}}{\mathbb{E}\{\alpha\}}  \bigg\|\\
    &\le \bigg\| \frac{\mathcal{A}}{\alpha}  \bigg\| \bigg| 1 - \frac{\alpha}{\mathbb{E}\{\alpha\}} \bigg| + \bigg\| \frac{\mathcal{A} - \mathbb{E}\{\mathcal{A}\}}{\mathbb{E}\{\alpha\}}  \bigg\|.
\end{align*}
We first bound terms related to $\alpha$. Setting $\delta = 1/2$ and we obtain from Lemma~\ref{lem_chernoff_pf_thm_state_concentration} that 
\begin{align*}
    &\PP\{\alpha - \EE\{\alpha\} \le -\EE\{\alpha\}/2 \} \\
    &\le 2 \exp\{ - \EE\{\alpha\}/12 \} \\
    &\sim 2 \exp \{- [(\nu_1^2+\nu_2^2) p_s + 2 \nu_1 \nu_2 p_d] n^2 / 24 \}.
\end{align*}
Utilizing Lemma~\ref{lem_chernoff_pf_thm_state_concentration} again with $\delta = 1/\sqrt{n}$, we have that
\begin{align*}
    &\PP\{ |\alpha - \EE\{\alpha\}| \ge \EE\{\alpha\}/\sqrt{n} \}\\
    &=
    \PP \bigg\{\bigg| \frac{\alpha}{\EE\{\alpha\}} - 1 \bigg| \ge \frac{1}{\sqrt{n}} \bigg\}\\
    &\le 
    2 e^{-\EE\{\alpha\}/(3n)} \\
    &\sim 
    2 n^{-[(\nu_1^2+\nu_2^2) p_s + 2 \nu_1 \nu_2 p_d]n / (6 \log n)}.
\end{align*}
Now we decompose $\mathcal{A} - \mathbb{E}\{\mathcal{A}\}$ as the sum of independent random matrices $\sum_{i=1}^n \sum_{j=i+1}^n Y_{ij}$ where $Y_{ij} := (a_{ij} - \mathbb{E}\{a_{ij}\}) (E_{ij} + E_{ji})$. Thus, $\|Y_{ij}\| \le 2$ and $\textup{var}(Y_{ij}) = \mathbb{E}\{Y_{ij}^2\} = (\mathbb{E}\{a_{ij}\} - (\mathbb{E}\{a_{ij}\})^2) (E_{ii} + E_{jj})$, where $E_{ij} = \mathbf{e}_i \mathbf{e}_j^T$. Hence 
\begin{align*}
v^2 &= \bigg\|\sum_{i=1}^n\sum_{j=i+1}^n \textup{var}(Y_{ij}) \bigg\| \\
&\le n \max\{\nu_1 p_s + \nu_2 p_d, \nu_1 p_d + \nu_2 p_s\} \\
&=: \beta_1 .
\end{align*} 
Therefore, from Lemma~\ref{lem_bernstein_pf_thm_state_concentration}, it follows that
\begin{align*}
    \PP\{\|\mathcal{A} - \EE\{\mathcal{A}\}\| > a\} \le 2n \exp\bigg\{\frac{-a^2}{2\beta_1  + 4a/3} \bigg\}.
\end{align*}
Let $a = C_1 \sqrt{\beta_1 \log n}$ for some constant $C_1$, and it holds that
\begin{align*}
    &\PP\{\|\mathcal{A} - \EE\{\mathcal{A}\}\| > C_1 \sqrt{\beta_1 \log n}\} \\
    &\le 2n \exp\bigg\{\frac{-C_1^2 \beta_1 \log n}{2\beta_1 + 4C_1\sqrt{\beta_1 \log n}/3} \bigg\}\\
    &= 2 n^{1 - C_1^2/(2 + 4C_1 \sqrt{\log n/\beta_1}/3)}.
\end{align*}
The right-hand side of the preceding equation tends to zero as $n \to \infty$ when the constant $C_1$ is large enough and $\log n / n = O(\max\{p_s,p_d\})$.

Finally, we need a bound for $\|\mathcal{A}\|$. Note that $\|\mathcal{A}\| \le \|\mathbb{E}\{\mathcal{A}\}\| + \|\mathcal{A}- \mathbb{E}\{\mathcal{A}\}\|$. Since we have already bound the second term, it suffices to study the spectrum of $\mathbb{E}\{\mathcal{A}\}$. Observe that $\mathbb{E}\{\mathcal{A}\} + p_s I_n$ has rank $2$ and its two eigenvectors corresponding to the nonzero eigenvalues are $[\bfl_{\nu_1n}^T~\bfl_{\nu_2n}^T]^T$ and $[\bfl_{\nu_1n}^T~-\bfl_{\nu_2n}^T]^T$. We can compute nonzero eigenvalues and obtain the upper bound of their absolute values as follows
\begin{align*}
    &\frac{n}{2} \Big| p_s \pm \sqrt{(1 - 4 \nu_1 \nu_2) p_s^2 + 4 \nu_1 \nu_2 p_d^2} \Big|\\
    &\le \frac{n}{2} \Big(p_s + \sqrt{(1 - 4\nu_1\nu_2)p_s^2 + 4 \nu_1 \nu_2 p_d^2} \Big) =: \beta_2.
\end{align*}
So $\|\mathbb{E}\{\mathcal{A}\}\| \le \beta_2 + p_s$. To sum up, it holds that
\begin{align*}
    &\bigg \|\tilde{W} - \frac{\mathbb{E}\{\mathcal{A}\}}{\mathbb{E}\{\alpha\}}  \bigg\|\\
    &\le 
    \bigg\| \frac{\mathcal{A}}{\alpha}  \bigg\| \bigg| 1 - \frac{\alpha}{\mathbb{E}\{\alpha\}}\bigg| + \bigg\| \frac{\mathcal{A} - \mathbb{E}\{\mathcal{A}\}}{\mathbb{E}\{\alpha\}} \bigg\|\\
    &\le 
    \frac{2(C_1 \sqrt{\beta_1 \log n} + \beta_2 + p_s)}{\mathbb{E}\{\alpha\}\sqrt{n}}  + \frac{C_1 \sqrt{\beta_1 \log n}}{\mathbb{E}\{\alpha\}} \\
    &\le \frac{C}{n},
\end{align*}
with probability 
\begin{align*}
    &1 - 2 \exp \{- [(\nu_1^2+\nu_2^2) p_s + 2 \nu_1 \nu_2 p_d] n^2 / 24 \} \\
    &- 2 n^{-[(\nu_1^2+\nu_2^2) p_s + 2 \nu_1 \nu_2 p_d] n / (6 \log n)} \\
    &- 4 n^{1 - C_1^2/(2 + 4C_1\sqrt{\log n/\beta_1}/3)} \to 1,
\end{align*}
as $n\to \infty$, if $\log n/n = O(\min\{p_s,p_d\})$. \hfill$\Box$
}

\section{Result on multiple-community case}\label{append_multiple case}
{\color{black}

In this section, we provide a result on the expression of $(I-\bar{A})^{-1}\bar{B}$ in the multiple-community case without proof, as a counterpart of Proposition~\ref{thm_expectation_structure}, and briefly discuss how to generalize Algorithm~\ref{alg_1} to the multiple-community case.

\subsection{Notation}
% empty set $\emptyset$

Let $[n] := \{1,2,\dots,n\}$, where $n$ is a positive integer. Denote the set consisting of all ordered $d$-tuples from the set $[n]$ by $[n]^d_<$, namely,
\begin{align*}
    [n]^d_< := \{(i_1, \dots, i_d) \in [n]^d : 1\le i_1 < i_2 < \cdots < i_d \le n\},
\end{align*}
where $[n]^d$ is the $d$-fold Cartesian product of $[n]$. More generally, for a set $\mathcal{S}$ containing finite number of distinct positive integers, we denote $\prescript{}{-i}{\mathcal{S}} := \mathcal{S}\setminus\{i\}$ for $i \in \mathcal{S}$, and use $\mathcal{S}^d_<$ to represent the set consisting of all ordered $d$-tuples from the set $\mathcal{S}$, i.e., 
\begin{align*}
    &\mathcal{S}^d_< := \{(i_1, \dots, i_d) \in \mathcal{S}^d : \\
    &\qquad \min \mathcal{S} \le i_1 < i_2 < \cdots < i_d \le \max \mathcal{S}\},
\end{align*}
where $\min \mathcal{S}$ and $\max \mathcal{S}$ are the minimum and maximum of $\mathcal{S}$. Specifically, let $\prescript{}{-i}{[n]} := [n] \setminus \{i\}$ for $i \in [n]$, and $\prescript{}{-i}{[n]}^d_<$ stands for the set consisting of all ordered $d$-tuples from~$\prescript{}{-i}{[n]}$.

\subsection{Result and discussion}
We consider the scenario where the agents can be partitioned into multiple groups $\mathcal{V}_1, \dots, \mathcal{V}_K$ (i.e., $\mathcal{V} = \cup_{k=1}^K \mathcal{V}_k$ and $\mathcal{V}_k \cap \mathcal{V}_l = \emptyset$ for all $k,l\in [K]$). To ease notation, we assume that $\mathcal{V}_1 = \{1, \dots, n_1\}$, $\mathcal{V}_2 = \{n_1 + 1, \dots, n_1 + n_2\}$, $\dots$, $\mathcal{V}_K = \{1 + \sum_{i=1}^{K-1} n_i, \dots, n_K + \sum_{i=1}^{K-1} n_i\}$ with $n_i := |\mathcal{V}_i|$, $i \in [K]$, and $\sum_{i=1}^{K} n_k = n$. Also, sort regular and stubborn agents in each community in the following way, $\mathcal{V}_{ri} = \{1 + \sum_{j=1}^{i-1} n_j, \dots, n_{ri} + \sum_{j=1}^{i-1} n_j\}$, $\mathcal{V}_{si} = \{1 + n_{ri} + \sum_{j=1}^{i-1} n_j, \dots, n_{i} + \sum_{j=1}^{i-1} n_j\}$, $i \in [K]$. Here, $\mathcal{V}_{ri}$ (resp. $\mathcal{V}_{si}$) is the set of regular (resp. stubborn) agents in community $i$. Let $n_{ri} := |\mathcal{V}_{ri}|$ and $n_{si} := |\mathcal{V}_{si}|$. Finally, denote the cardinality of regular and stubborn agents by $n_r := |\mathcal{V}_r| = \sum_{i=1}^K n_{ri}$ and $n_s := |\mathcal{V}_s| = \sum_{i=1}^K n_{si}$ respectively. Assumptions in the multiple-community case, similar to Assumption~\ref{asmp_community}, are summarized as follows. 

\begin{assum}\label{asmp_community_multiple}~\\
(1.a) The agent set $\mathcal{V}$ consists of $K$ communities with $K \ge 2$, $n_i > 0$, $i \in [K]$, and $\sum_{i=1}^K n_i = n$. \\
(1.b) All communities have regular agents, namely, $1 \le n_{ri} \le n_i$, $i \in [K]$.\\
(2) The within-group and inter-group interaction probabilities are $w_s, w_d > 0$, respectively, with $w_s \not = w_d$, and \[\left(\sum_{i=1}^K n_i(n_i-1) \right) w_s + \left( \sum_{i=1}^K \sum_{j\not= i} n_i n_j \right) w_d = 2.\]
(3) $X(0)$ is deterministic. It holds that $X^r(0) \in \mathcal{S}$, with
\begin{align*} 
\mathcal{S} := \{x \in \mathbb{R}^{n_r} : x_i \in [\underline{s}, \overline{s}], 1\le i \le n_r\},
\end{align*}
where $\underline{s} := \min_{1\le i\le n_s}\{\mathbf{x}_i^s\}$, $\overline{s}:= \max_{1\le i\le n_s} \{\mathbf{x}_i^s\}$, $\mathbf{x}^s := X^s(0) = [(\mathbf{x}^{s1})^T ~ (\mathbf{x}^{s2})^T ~ \cdots ~ (\mathbf{x}^{sK})^T]^T$ is  the stubborn state vector, and $\mathbf{x}^{sk}$ is the vector for the community~$k$, $k \in [K]$. 
\end{assum}

The following proposition generalizes Proposition~\ref{thm_expectation_structure} to the multiple-community case under Assumption~\ref{asmp_community_multiple}. The proof directly follows from the inverse formula of block matrices~\cite{henderson1981deriving} and is omitted. 

\begin{prop}\label{prop_inverse_block_multiple}
Suppose that Assumption~\ref{asmp_community_multiple} holds and there exists at least one stubborn agent in the network (i.e., $n_r < n$). Then $(I-\bar{A})^{-1}$ exists, and it holds that
\begin{align*}%\label{eq_block_I-A}
    (I-\bar{A})^{-1} = (1-q)^{-1} \begin{bmatrix}
    \tilde{A}^{(11)} & \tilde{A}^{(12)} & \cdots & \tilde{A}^{(1K)}\\
    \tilde{A}^{(21)} & \ddots & \cdots & \vdots \\
    \vdots & \cdots & \ddots & \tilde{A}^{(K-1,K)} \\
    \tilde{A}^{(K1)} & \cdots & \tilde{A}^{(K,K-1)} & \tilde{A}^{(KK)}
    \end{bmatrix},
\end{align*}
where 
\begin{align*}
    &\tilde{A}^{(ii)} = \frac{1}{w_sn_i +  w_d  \sum_{j\not= i} n_j } \\
    & \bigg[I_{n_{ri}} - \frac{1}{n_{ri}} \bigg( 1 - \frac{d_A(\prescript{}{-i}{[K]})}{d_A([K])} \bigg) \bfl_{n_{ri},n_{ri}} \bigg], ~ i \in [K],\\
    &\tilde{A}^{(ij)}  \\
    &=\frac{w_d e_A([K]\setminus\{i,j\}) \bfl_{n_{ri},n_{rj}}}{(w_sn_i +  w_d  \sum_{k\not= i} n_k ) (w_sn_j +  w_d  \sum_{k\not= j} n_k ) d_A([K])}, \\ & ~ i,j \in [K],~i\not=j.
\end{align*}
Here
\begin{align*} \nonumber
    d_A(\mathcal{S}) &= - \sum_{p=0}^{|\mathcal{S}|} \sum_{(j_1,\dots,j_p) \in \mathcal{S}^p_<} \bigg[ (w_d - w_s)^{p-1} \\%\label{eq_block_I-A_dS}
    &(w_s + (p-1) w_d) \prod_{l=1}^p \bigg(\frac{n_{r,j_l}}{w_s n_{j_l} + w_d \sum_{k\not=j_l} n_k}\bigg) \bigg],\\ %\label{eq_Block_I-A_eS}
    e_A(\mathcal{S}) &= \prod_{p \in \mathcal{S}} \bigg(1 + \frac{(w_d-w_s) n_{rp}}{w_sn_p +  w_d \sum_{j\not= p} n_j} \bigg),
    % &= \sum_{i=0}^{|\mathcal{S}|} \sum_{(j_1,\dots,j_i) \in \mathcal{S}_<^i} \bigg[ (w_d-w_s)^i \prod_{l=1}^i \bigg(\frac{n_{r_{j_l}}}{w_s n_{j_l} + w_d \sum_{k\not=j_l} n_k}\bigg) \bigg],
\end{align*}
where $\mathcal{S}$ is a set consisting of finite number of distinct positive integers, for $p=0$ we define the term $\sum_{(j_1,\dots,j_p) \in \mathcal{S}^p_<}  \prod_{l=1}^p (n_{r,j_l}/(w_s n_{j_l} + w_d \sum_{k\not=j_l} n_k)) = 1$, and $d_A(\emptyset) = e_A(\emptyset) = 1$. In addition,
\begin{align*}
    (I-\bar{A})^{-1}\bar{B} = 
    \begin{bmatrix}
    \tilde{B}^{(11)} & \tilde{B}^{(12)} & \cdots & \tilde{B}^{1K}\\
    \tilde{B}^{(21)} & \ddots & \cdots & \vdots \\
    \vdots & \cdots & \ddots & \tilde{B}^{(K-1,K)} \\
    \tilde{B}^{(K1)} & \cdots & \tilde{B}^{(K,K-1)} & \tilde{B}^{(KK)}
    \end{bmatrix},
\end{align*}
where
\begin{align*}
    \tilde{B}^{ii} &= \frac{1}{n_{ri}} \bigg( \frac{d_A(\prescript{}{-i}{[K]})}{d_A([K])} - 1 \bigg) \bfl_{n_{ri},n_{si}},~ i\in[K],\\
    \tilde{B}^{ij} &= \frac{w_d e_A([K]\setminus\{i,j\})}{(w_sn_i +  w_d  \sum_{k\not= i} n_k)d_A([K])} \bfl_{n_{ri},n_{sj}},\\
    &i,j \in [K],~ i\not=j.
\end{align*}
\end{prop}

\begin{rem}
	The above proposition provides a parallel result of Proposition~\ref{thm_expectation_structure}, and shows that $(I-\bar{A})^{-1}\bar{B}$ has a block structure and hence so does $\mathbf{x}^r$. The developed framework in the current paper indicates that studying a condition similar to Assumption~\ref{asmp_stubborn_agents} would be the key to addressing the problem in the general case, but such condition would be more complex and the investigation of it is beyond the scope of this paper. Note that it is possible to generalize Algorithm~\ref{alg_1} to the multiple-community case by examining the structure of $\mathbf{x}^r$ more carefully, and the techniques we developed in the current paper would contribute to the analysis of the generalized algorithm.
\end{rem}

}

\section{Some results on stochastic approximation}\label{append_SA}

In this section we introduce some results, which are used throughout the later appendices, on a linear stochastic approximation algorithm. Consider deterministic matrices $H(t), H \in \mathbb{R}^{l\times l}$ and random vectors $z(t), e(t), v(t) \in \mathbb{R}^l$, and define the following linear recursion, 
\begin{align}\label{eq_linearRecursion}
    z(t+1) &= z(t) + a(t) H(t) z(t) + a(t) (e(t) + v(t))
\end{align}
The following results follow from Lemma~3.1.1 and Theorem~3.1.1 of \cite{chen2002stochastic}.

\begin{prop}\label{prop_append_SA}
Assume $a(t)$ is such that $a(t) > 0$, $a(t) \to 0$ as $t \to \infty$, $\sum_{t=1}^{\infty} a(t) = \infty$, and
\begin{align}\label{eq_alpha}
    a^{-1}(t+1) - a^{-1}(t) \to \alpha \ge 0.
\end{align}
If $\sum_{t=1}^{\infty} a(t)^{1-d} e(t) < \infty$ and $v(t) = o(a(t)^d)$ as $t \to \infty$ for some $d\in[0,1/2)$, $H(t) \to H$, and $H + \alpha d I$ is stable (all its eigenvalues are with negative real parts), then $z(t)/a(t)^d \to 0$.
\end{prop}

%\begin{pf}
%The first conclusion is obtained from Lemma~3.1.1 and Theorem~3.1.1 of \cite{chen2002stochastic}. For the second conclusion, from~(3.3.38) in the proof of Theorem~3.3.1 of \cite{chen2002stochastic}, it suffices to prove that $\sum_{i=1}^t \xi(t,i) \overset{\textup{d}}{\to} \mathcal{N}(\mathbf{0},\mathbf{S})$.
%
%Assumption~(ii) and the definition of $\xi(t,i)$ imply that $\{\sum_{i=1}^k \xi(t,i), \mathcal{F}(k), 1\le k \le t, t \in \mathbb{N}^+\}$ is a zero-mean, square-integrable martingale array. Hence the conclusion follows from Theorem~3.2 of \cite{hall2014martingale}, combined with Cram{\'e}r-Wold device, i.e., that for a sequence of random vectors $\{X_n\}$ in $\mathbb{R}^n$, $X_n \overset{\textup{d}}{\to} X$ if and only if $x^T X_n \overset{\textup{d}}{\to} x^T X$ for all $x \in \mathbb{R}^n$. \hfill$\Box$
%\end{pf}

\section{Proof of Theorem \ref{thm_stability}}\label{append_proof_thm_stability}    % Each appendix must have a short title.

From Theorem~2.1 of \cite{diaconis1999iterated} we have the following stability result for Markov chains.

\begin{prop}\label{lem3} Consider a Markov chain on $\mathbb{R}^n$, defined by 
\[
Y(t + 1) = A(t) Y(t) + U(t), \quad t \in \mathbb{N},
\]
where $\{[A(t)~ U(t)], t \in \mathbb{N}\}$ is a sequence of i.i.d. random matrices taking values in~$\mathbb{R}^{n \times (n+1)}$, such that $\mathbb{E} \{\log^+ \|A(t)\| \} < \infty$ and $\mathbb{E} \{\log^+ \|U(t)\| \} < \infty$ ($x^+ = x$ if $x > 0$, $x^+ = 0$ if $x \le 0$). Suppose that
\[
\inf_{t > 0} \frac1t \mathbb{E} \left\{\log \Big\|\overleftarrow{\Phi}_A(0,t-1)\Big\| \right\} < 0,
\]
where $\overleftarrow{\Phi}_A(s, t) := A(s) \cdots A(t)$ for $0 \le s \le t$,~and $\overleftarrow{\Phi}_A(s, t) = I$ for $s > t$. In addition, the only~invariant subspace of $\mathbb{R}^n$ is itself, where an invariant subspace is a linear subspace $L \subset \mathbb{R}^n$ such that $\mathbb{P}\{Y(1) \in L|Y(0) = y\} = 1$ for all $y \in L$. Then the infinite random series $Y^* = \sum_{j = 0}^{\infty} \overleftarrow{\Phi}_A(0, j - 1)$ $U(j)$ converges a.s., and its distribution is the unique stationary distribution of the Markov chain $\{Y(t)\}$.
\end{prop}

We also need the following lemma for substochastic matrices to establish Corollary~\ref{cor_rho_barA}, which will be used in the proof of Theorem~\ref{thm_stability}. 
\begin{lem}\label{lem_substochastic}
Consider a substochastic matrix $A = [a_{ij}] \in \mathbb{R}^{n \times n}$. If for every row $i$, $1 \le i \le n$, there exists an integer~$j, ~ 1 \le j \le n$, satisfying that the sum of $j$-th row less than one, and a sequence of distinct integers $k_1 = i, ~ k_2, \dots, ~ k_m =~j, ~ 1 \le m \le n$, such that $a_{k_1 k_2} a_{k_2 k_3} \cdots a_{k_{m - 1} k_m} > 0$, then $\rho(A) < 1$. 
\end{lem}

\begin{pf}
From Theorem~8.3.1 of \cite{horn2012matrix}, we know that there is a nonnegative nonzero vector $x \in \mathbb{R}^n$ such that 
\begin{align*}
    x^T A = \rho(A) x^T.
\end{align*}
Let $\mathcal{T} := \{1 \le i \le n: x_i > 0\}$, then
\begin{align*}
    \rho(A) \sum\nolimits_{i\in \mathcal{T}} x_i &= \sum\nolimits_{i \in \mathcal{T}} (A^T x)_i\\
    &= \sum\nolimits_{i \in \mathcal{T}} \Big(\sum\nolimits_{1 \le j \le n} a_{ji} x_j\Big)\\
    &= \sum\nolimits_{1\le j \le n} x_j \sum\nolimits_{i\in\mathcal{T}} a_{ji}\\
    &= \sum\nolimits_{j \in \mathcal{T}} x_j \sum\nolimits_{i\in\mathcal{T}} a_{ji}.
\end{align*}
There must exists $j \in \mathcal{T}$ such that $\sum_{i \in \mathcal{T}} a_{ji} < 1$. Otherwise, $\sum_{i \in \mathcal{T}} a_{ji} = 1$ for all $j \in \mathcal{T}$ and $\mathcal{T} \subsetneqq \{1,\dots,n\}$ by assumption, which means that $a_{jk} = 0$, for all $k \not\in \mathcal{T}$. This contradicts with the assumption. So $\rho(A)\sum_{i \in \mathcal{T}} x_i < \sum_{j \in \mathcal{T}} x_j$, and $\rho(A) < 1$.\hfill$\Box$
\end{pf}

\emph{\textbf{Proof of Theorem \ref{thm_stability}:}}

We use Proposition~\ref{lem3} and Corollary~\ref{cor_rho_barA} to verify the first part of (i). 
Note that
\begin{align}\nonumber
&\mathbb{E} \Big\{ \Big\| \overleftarrow{\Phi}_A(0,t) \Big\|_{1} \Big\} \\\nonumber
&= 
\mathbb{E} \Big\{ \underset{1 \le j \le n}{\text{max}} \sum\nolimits_{1 \le i \le n} \Big|\Big[ \overleftarrow{\Phi}_A(0,t) \Big]_{ij}\Big| \Big\} \\\nonumber
&\le 
\sum\nolimits_{1 \le j \le n} \sum\nolimits_{1 \le i \le n} \mathbb{E} \Big\{ \Big|\Big[ \overleftarrow{\Phi}_A(0,t) \Big]_{ij}\Big| \Big\} \\\nonumber
&\le
n \Big\| \mathbb{E} \Big\{ \overleftarrow{\Phi}_A(0,t) \Big\} \Big\|_{\infty} \\\label{eq_keyToCoupling}
&\le 
\gamma n (t+1)^{n - 1} \rho(\bar{A})^{t+1},
\end{align}
for some constant $\gamma > 0$, where the last equality follows from the Jordan canonical decomposition. Thus from Jensen's inequality, we have that
\begin{align*}
    &\inf_{t > 0} \frac1t \mathbb{E} \Big\{\log \Big\|\overleftarrow{\Phi}_A(0,t-1)\Big\|_1 \Big\}\\
    &\le \lim_{t \to \infty} \frac{\log (\gamma n t^{n - 1} \rho(\bar{A})^t)}{t} = \log \rho(\bar{A}) < 0,
\end{align*}
from Corollary \ref{cor_rho_barA}.

Assumption~(iii), the assumption for the existence of stubborn agents, and update rule~\eqref{eq_update_rule1} ensure that $\mathbb{R}^n$ is the only invariant subspace of itself. So let $U(t) := B(t)X^s(t)$, and it follows from Proposition \ref{lem3} that $X^r_* := \sum_{j = 0}^{\infty} \overleftarrow{\Phi}_A(0,j - 1) U(j)$ converges a.s., and its distribution $\pi$ is the unique stationary distribution of $\{X^r(t)\}$. In addition, from Markov inequality and \eqref{eq_keyToCoupling},
\[\mathbb{P}\Big\{\Big\|\overleftarrow{\Phi}(0,t-1)\Big\|_1 \ge \varepsilon\Big\} \le \varepsilon^{-1}\gamma n t^{n - 1} \rho(\bar{A})^{t}.\]
So by Borel-Cantelli lemma, $\overleftarrow{\Phi}_A(0,t) X^r(0)$ converges to zero a.s. Hence $\overleftarrow{X}^r(t) := \overleftarrow{\Phi}_A(0,t-1) X^r(0) + \sum_{j = 0}^{t-1} \overleftarrow{\Phi}_A(0,j - 1) U(j)$ converges a.s. to $X^r_*$. Since $\overleftarrow{X}^r(t)$ and $X^r(t)$ has the same distribution, we have that $X^r(t) \overset{\textup{d}}{\to} \pi$, as $t \to \infty$.

For (ii) of the theorem, since $|U(t)| < L$ for some positive $L>0$, by dominated convergence theorem, $\mathbb{E}\{X^r_*\} = \lim_{t \to \infty} \mathbb{E} \{\overleftarrow{X}^r(t) \}$. It follows that $\mathbb{E}\{ X^r_*\} = \sum_{j = 0}^{\infty} \bar{A}^j \bar{u} = (I - \bar{A})^{-1} \bar{u}$ from independence and $\rho(\bar{A}) < 1$, where $\bar{u} := \mathbb{E}\{U(t)\} = \bar{B}\textbf{x}^s$. Finally, $X^r(t)$ and $\overleftarrow{X}^r(t)$ have the same distribution, so $\mathbb{E}\{X^r(t)\} = \mathbb{E}\{\overleftarrow{X}^r(t)\}$, and \eqref{eq_expectation_limit} holds.

To prove (iii), write
\begin{align*}
    S^r(t+1) = S^r(t) - \frac{1}{t+1} S^r(t) + \frac{1}{t+1} X^r(t).
\end{align*}
Using notations in Appendix~\ref{append_SA}, let $z(t) = S^r(t) - \mathbf{x}^r$, $a(t) = 1/(t+1)$, $H(t) \equiv H = -I$, $e(t) = X^r(t) - \mathbb{E}\{X^r(t)\}$, and $v(t) = \mathbb{E}\{X^r(t)\} - \mathbf{x}^r$. So from Proposition~\ref{prop_append_SA} in Appendix~\ref{append_SA}, to show $z(t) \to 0$, it suffices to validate that $\sum_{t=1}^{\infty} a(t) e(t) < \infty$ and $v(t) \to 0$ as $t \to \infty$. The latter follows from (ii). Note that
\begin{align}\nonumber
    &\sum\nolimits_{t=1}^{\infty} a(t) e(t)\\\nonumber
    &=
    \sum\nolimits_{t=1}^{\infty} a(t) (X^r(t) - \mathbb{E}\{X^r(t)\})\\\nonumber
    &=
    \sum\nolimits_{t=1}^{\infty} a(t) (A(t-1) X^r(t-1) + U(t-1) \\\nonumber
    &~  - \bar{A} \mathbb{E}\{X^r(t-1)\} - \bar{u})\\\nonumber
    &= \sum\nolimits_{t=1}^{\infty} a(t) (A(t-1) - \bar{A}) X^r(t-1)\\\nonumber
    &~ + \sum\nolimits_{t=1}^{\infty} a(t) (U(t-1) - \bar{u}) \\\nonumber
    &~ + \bar{A} \sum\nolimits_{t=1}^{\infty} a(t) (X^r(t-1)-\mathbb{E}\{X^r(t-1)\})\\\nonumber
    &=
    \sum\nolimits_{t=1}^{\infty} a(t) (A(t-1) - \bar{A}) X^r(t-1)\\\nonumber
    &~ + \sum\nolimits_{t=1}^{\infty} a(t) (U(t-1) - \bar{u}) \\\nonumber
    &~ + \bar{A} \sum\nolimits_{t=0}^{\infty} a(t) (X^r(t)-\mathbb{E}\{X^r(t)\})\\\nonumber
    &~ + \bar{A} \sum\nolimits_{t=1}^{\infty} (a(t) - a(t-1))\\\label{eq_pf_thm_stability_noise_series}
    &\qquad\qquad (X^r(t-1)-\mathbb{E}\{X^r(t-1)\})
\end{align}
On the right side of the above equality, the first two terms are weighted sums of martingale difference sequences, and converge by Theorem~B.6.1 of \cite{chen2002stochastic}, and the last term converges since $X^r(t)$ is bounded and $a(t) - a(t-1)$ $ = -1/t(t+1)$. Hence $\sum_{t=1}^{\infty} a(t) e(t) < \infty$ is obtained by combining the left side and the third term on the right together, and noting that $(I - \bar{A})^{-1}$ exists by Corollary~\ref{cor_rho_barA}.~\hfill$\Box$

\begin{rem}\label{rmk_thm_stability_convergence_rate}
Note that, similar to the proof of Theorem~\ref{thm_stability}~(iii), we are able to show that $\sum_{t=1}^{\infty} a(t)^{1-d} e(t) < \infty$, for all $d\in [0,1/2)$. Since $\mathbb{E}\{X^r(t)\}$ converges exponentially to $\mathbf{x}^r$, it holds that $v(t) = o(a(t)^d)$ for all $d \in [0,1/2)$. Therefore, from Proposition~\ref{prop_append_SA}, it follows that 
$(S^r(t) - \mathbf{x}^r)/a(t)^d \to 0$ a.s., for all $d\in [0,1/2)$.
\end{rem}

\section{Proof of Theorem \ref{thm_convergence}}\label{append_proof_thm_convergence} 

We know from \eqref{eq_ergodicity} that $S^r(t) \to \mathbf{x}^r$ a.s., as $t \to \infty$, where $S^r(t) = [S^r_i(t)]_{1\le i \le n_r}$. That is, $S^r_i(t) \to \chi_1$ for $1\le i \le n_{r1}$ and $S^r_i(t) \to \chi_2$ for $n_{r1}+1 \le i \le n_r$. Hence for $\varepsilon = |\chi_1 - \chi_2|/(2n_r(n_r+1))$, there exists an integer-valued random variable $T$ such that for all $t > T$,
$|S^r_i(t) - \chi_1| < \varepsilon$ for $1\le i \le n_{r1}$, and $|S^r_i(t) - \chi_2| < \varepsilon$ for $n_{r1}+1 \le i \le n_r$. Since Assumption \ref{asmp_stubborn_agents} ensures that $\chi_1 \not= \chi_2$, we can assume that $\chi_1 > \chi_2$. Consequently, from $n_{r1} \ge 1$ and $n_{r2} \ge 1$,
\begin{align*}
    (\chi_1 - \varepsilon) - \Big(\frac{n_{r1}\chi_1 + n_{r2}\chi_2}{n_r} + n_r\varepsilon \Big) &> 0,\\
    (\chi_2 + \varepsilon) - \Big(\frac{n_{r1}\chi_1 + n_{r2}\chi_2}{n_r} - n_r\varepsilon \Big) &< 0.
\end{align*}
This fact means that $S^r_i(t) > \bar{s}^r(t)$ for $1\le i \le n_{r1}$ and $S^r_i(t) < \bar{s}^r(t)$ for $n_{r1}+1 \le i \le n_r$, $\forall t > T$, which implies the finite-time convergence of the community recovery step, combined with Assumption \ref{asmp_stubborn_agent_structure}.

Now we can assume that the true community has been recovered since this recovery step converges in finite time~$T$. As a consequence, we know the community structure of all agents for $t > T$, and also the size of both communities (i.e., $n_1$ and $n_2$). In other words, $\hat{n}_k(t) = n_k$, $k = 1,2$, $t > T$. Hence $\hat{w}_s(t)$ in Line $5$ of Algorithm \ref{alg_1} can be rewritten as follows for $t > T+1$,
\begin{equation}\label{eq_parameter_estimation_recursion}
    \hat{w}_s(t) = \hat{w}_s(t-1) -  \frac{a}{t} \text{sgn}(g(t)) \Big(g(t) \hat{w}_s(t-1) + \frac{h_2(t)}{n_1n_2}\Big),
\end{equation}
where $g(t) = h_1(t) - \frac{n_1^2+ n_2^2 - n_1 - n_2}{2n_1n_2} h_2(t)$ and 
\begin{align*}
    h_1(t) &= \frac{n_{s1}}{n_{r1}} \sum\nolimits_{i\in \mathcal{V}_{r1}} S^r_i(t) - \mathbf{1}_{n_{s1}}^T \mathbf{x}^{s1}, \\
    h_2(t) &= \frac{n_2}{n_{r1}} \sum\nolimits_{i\in \mathcal{V}_{r1}} S^r_i(t) - \sum\nolimits_{i\in \mathcal{V}_{r2}} S^r_i(t) - \mathbf{1}_{n_{s2}}^T \mathbf{x}^{s2}.
\end{align*}
In order to utilize Proposition~\ref{prop_append_SA} in Appendix~\ref{append_SA}, let $z(t) = \hat{w}_s(t) - w_s$, $a(t) = a/(t+1)$, $H(t)=-|g(t+1)|$, and \[v(t) = -\text{sgn}(g(t+1))\Big(\frac{h_2(t+1)}{n_1n_2} + g(t+1)w_s\Big).\] 
From the fact that \eqref{eq_linear_system} has a unique solution, it holds that $H(t) \to H := - |\eta| < 0$, where~$\eta = $ $h_1^* - \frac{n_1^2 + n_2^2 - n_1 - n_2}{2n_1n_2} h_2^* \not= 0$ with $h_1^* = n_{s1} \chi_1 - \mathbf{1}_{n_{s1}}^T \mathbf{x}^{s1}$ and $h_2^* = n_2 \chi_1 - n_{r2}\chi_2 - \mathbf{1}_{n_{s2}}^T \mathbf{x}^{s2}$. Notice that 
\begin{align*}
    &\frac{h_2(t)}{n_1n_2} + g(t)w_s\\
    &=
    h_1(t) w_s + \frac{h_2(t)}{2n_1n_2}(2 - (n_1^2+n_2^2 - n_1 - n_2)w_s)\\
    &\to
    h_1^* w_s + \frac{h_2^*}{2n_1n_2}(2 - (n_1^2+n_2^2 - n_1 - n_2)w_s), 
\end{align*}
from \eqref{eq_ergodicity}, and this limit is $h_1^*w_s + h_2^*w_d = 0$ by~\eqref{eq_linear_system}, implying $v(t) \to 0$. So the interaction estimator converges by Proposition~\ref{prop_append_SA} with $d=0$.\hfill$\Box$

\section{Proof of Lemma~\ref{lem_concentration}}\label{append_proof_lem_concentration}

In this section, we prove Lemma~\ref{lem_concentration} by leveraging the techniques introduced in~\cite{glynn2002hoeffding}. We need the following result.

\begin{lem}[Lemma~8.1 of~\cite{devroye2013probabilistic}]\label{lem_hoeffding}~\\
	Let $X$ be a random variable with $\mathbb{E}\{X\} = 0$ and $c_1\le X\le c_2$ for some constants $c_1$ and $c_2$. Then for all $\lambda > 0$, $\mathbb{E}\{\exp\{\lambda X\} \} \le \exp\{\lambda^2 (c_2-c_1)^2/8\}$.
\end{lem}

In the proof, we denote $\mathbb{P}_x\{\cdot\}:=\mathbb{P}\{\cdot|X(0)=x\}$ and $\mathbb{E}_x\{\cdot\}:=\mathbb{E}\{\cdot|X(0)=x\}$ to ease notation.

 Let $g(X(i)) := \sum\nolimits_{t=i}^{\infty} \mathbb{E}\{f(X(t)) - \alpha | X(i)\}$, $i \in \mathbb{N}$, and it holds that 
\begin{align*}
	&g(x) - \mathbb{E}_x\{g(X(1))\} \\
	&= \sum_{t=0}^{\infty} \mathbb{E}_x\{f(X(t)) - \alpha\} - \sum_{t=1}^{\infty} \mathbb{E}_x\{f(X(t)) - \alpha | X(1)\} \\
	&= \mathbb{E}_x\{f(X(0)) - \alpha\} + \sum_{t=1}^{\infty} \mathbb{E}_x\{f(X(t)) - \alpha\} \\
	&\quad - \sum_{t=1}^{\infty} \mathbb{E}_x\{f(X(t)) - \alpha | X(1)\} \\
	&= \mathbb{E}_x\{f(X(0)) - \alpha\}\\
	&=f(x) - \alpha.
\end{align*}
Hence, for $t \in \mathbb{N}$,
\begin{align*}
 	tS_f(t) - t\alpha 
 	&= \sum_{i=0}^{t-1} f(X(i)) - t \alpha\\
 	&= \sum_{i=0}^{t-1} g(X(i)) - \mathbb{E}\{g(X(i+1))|X(i)\}\\
	&=\sum_{i=1}^t D_i + g(X(0)) - g(X(t)),
\end{align*}
where $D_i := g(X(i)) - \mathbb{E}\{g(X(i))|X(i-1)\} = g(X(i)) - \mathbb{E}\{g(X(i))|X(0),\dots,X(i-1)\}$, $1\le i \le t$. It follows from $|g(x)| \le \|g\|_s$ for all $x\in \mathcal{X}$ that for $\lambda > 0$
\begin{align*}
	&\mathbb{E}\{\exp\{\lambda t(S_f(t) - \alpha)\}\}\\
	&\le 
	\exp\{2\lambda \|g\|_s\} \mathbb{E}\bigg\{\exp\bigg\{\lambda  \sum_{i=1}^t D_i\bigg\}\bigg\}\\
	&\le 
	\exp\{2\lambda \|g\|_s + t \lambda^2 \|g\|^2_s/2\},
\end{align*}
where in the last inequality we use Lemma~\ref{lem_hoeffding} for $D_i$ conditioned on $X(0)$, $\dots$, $X(i-1)$, $1\le i \le t$, inductively. Markov inequality yields that
\begin{align*}
	&\mathbb{P} \{S_f(t) - \alpha \ge \varepsilon \}\\
	&= 
	\mathbb{P}\{\exp\{\lambda t(S_f(t) - \alpha)\} \ge \exp\{\lambda t\varepsilon\} \}\\
	&\le
	\exp\{-\lambda t \varepsilon + 2\lambda \|g\|_s + t \lambda^2 \|g\|^2_s/2\}\\
	&\le 
	\exp\left\{-\frac{(t\varepsilon - 2 \|g\|_s)^2}{2 t \|g\|_s^2}\right\},
\end{align*}
where in the last inequality we let $\lambda = (t\varepsilon - 2 \|g\|_s)/t\|g\|_s^2$. Applying the same argument to $-S_f(t)$, we obtain the conclusion. \hfill$\Box$

\section{Proof of Theorem \ref{thm_finitesample}}\label{append_proof_thm_finitesample}

We use Lemma~\ref{lem_concentration} to proof the theorem. %the concentration inequality for Markov chains in \cite{glynn2002hoeffding}, but without the recurrent hypothesis therein, to show the conclusion. In fact, the proof of \cite{glynn2002hoeffding} indicates the following result.
%\begin{lem}
%Consider a Markov chain $\{X(t)\}$ taking values on a compact state space $\mathcal{X}$ and having a unique stationary distribution $\pi$. It holds for $S_f(t) := \frac1t \sum_{i=0}^{t-1}f(X(i))$, where $f: \mathcal{X} \to \mathbb{R}$, $\alpha := \int_{\mathcal{X}} f(x) \pi(dx)$, $\varepsilon > 0$, and $t > 2 \|g\|_s/\varepsilon$ that
%\begin{align*}
%    \mathbb{P}\{|S_f(t) - \alpha| \ge \varepsilon\} \le 2 \exp\left\{-\frac{(t\varepsilon - 2 \|g\|_s)^2}{2 t \|g\|_s^2}\right\},
%\end{align*}
%if $\|g\|_s < \infty$, where
%\begin{align*}
%    g(x) &:= \sum\nolimits_{t=0}^{\infty} \mathbb{E}\{f(X(t)) - \alpha | X(0) = x\},\\
%    \|g\|_s &:= \sup\{|g(x)|:x\in \mathcal{X}\}.
%\end{align*}
%\end{lem}
For the gossip model with stubborn agents, define $f_i(x) = x_i$, $x\in \mathbb{R}^{n_r}$, $1\le i \le n_r$, and we have that for $t > 2 \|g_i\|_s/\varepsilon$ and $\varepsilon > 0$
\begin{align*}
    \mathbb{P}\{|S^r_i(t) - \mathbf{x}^r_i| \ge \varepsilon\} \le 2 \exp\left\{-\frac{(t\varepsilon - 2 \|g_i\|_s)^2}{2 t \|g_i\|_s^2}\right\},
\end{align*}
where $g_i(x) := \sum_{t=0}^{\infty} \mathbb{E}\{X^r_i(t) - \mathbf{x}^r_i | X(0) = x\}$, $x \in \mathcal{S}$. Note that $g_i(x)$, $x\in \mathcal{S}$, is the $i$-th component of vector
\begin{align*}
    G(x) &:= \sum_{t=0}^{\infty} \mathbb{E}\{X^r(t) - \mathbf{x}^r | X(0) = x\}\\
    &=
    \sum_{t=0}^{\infty} \bigg(\bar{A}^t x + \sum_{i=0}^{t-1} \bar{A}^i \bar{B}\mathbf{x}^s - \sum_{i=0}^{\infty} \bar{A}^i \bar{B}\mathbf{x}^s \bigg)\\
    &=
    \sum_{t=0}^{\infty} \bigg(\bar{A}^t x - \sum_{i=t}^{\infty} \bar{A}^i \bar{B}\mathbf{x}^s \bigg),
\end{align*}
and
\begin{align*}
    \|G(x)\| &= \sum_{t=0}^{\infty} \bigg\| \bar{A}^t \bigg(x - \sum_{i=0}^{\infty} \bar{A}^i \bar{B}\mathbf{x}^s\bigg) \bigg\|\\
    &\le
    \sum_{t=0}^{\infty}  \rho^t(\bar{A}) \|x - \mathbf{x}^r\| \\
    &\le
    \frac{2 \sqrt{n_r} \max\{|\bar{s}|, |\underline{s}|\}}{1-\rho(\bar{A})} =: s_*,
\end{align*}
where the second inequality holds because $\bar{A}$ is symmetric, implying $\|\bar{A}\|_2 = \rho(\bar{A})$, and the last inequality follows from $x \in \mathcal{S}$. Since $|g_i(x)| \le \|G(x)\|$, we know that $\|g_i\|_s \le s_*$, and for $1\le i \le n_r$, $t > 2 s_*/\varepsilon$, and $\varepsilon > 0$, it holds that
\begin{align*}
    \mathbb{P}\{|S^r_i(t) - \mathbf{x}^r_i| \ge \varepsilon\} \le 2 \exp\left\{-\frac{(t\varepsilon - 2 s_*)^2}{2 t s_*^2}\right\}.
\end{align*}

Now let $\varepsilon_0 = |\chi_1 - \chi_2|/(2n_r(n_r+1))$, it follows from the proof of Theorem~\ref{thm_convergence} that
\begin{align}\nonumber
    &\mathbb{P}\{\hat{\mathcal{C}}(i,t) = \mathcal{C}(i), \forall i \in \mathcal{V}\}\\ \nonumber
    &\ge
    \mathbb{P}\{|S_i^r(t) - \mathbf{x}_i^r| < \varepsilon_0, \forall 1 \le i \le n_r\}\\\nonumber
    &\ge 
    1 - \sum_{i=1}^{n_r} \mathbb{P}\{|S_i^r(t) - \mathbf{x}_i^r| \ge \varepsilon_0\}\\
    &\ge
    1 - 2n_r \exp\left\{-\frac{(t\varepsilon_0 - 2 s_*)^2}{2 t s_*^2}\right\},
\end{align}
which leads to the conclusion, combined with the explicit form of $|\chi_1-\chi_2|$ given in the proof of Proposition~\ref{thm_identifiability}. \hfill$\Box$

\section{Proof of Theorem \ref{thm_convergencerate}}\label{append_proof_thm_convergencerate}

We follow the notations in the proof of Theorem~\ref{thm_convergence}. From~\eqref{eq_parameter_estimation_recursion}, it suffices to show that $z(t)/a(t)^d \to 0$ for some $d \in (0,1/2)$. From Remark~\ref{rmk_thm_stability_convergence_rate}, we know that $(S^r(t) - \mathbf{x}^r)/a(t)^d \to 0$ a.s., for all $d\in [0,1/2)$. Hence $(h_k(t) - h_k^*)/a(t)^d \to 0$, $k=1,2$, implying $v(t)/a(t)^d \to 0$ for all $d\in [0,1/2)$. Now note that $H + \alpha dI = -|\eta| + d/a$, where $\alpha$ is given in~\eqref{eq_alpha} of Proposition~\ref{prop_append_SA} and $a$ is defined in Algorithm~\ref{alg_1}. So it follows from Proposition~\ref{prop_append_SA} that $z(t)/a(t)^d \to 0$ for all $d \in [0, \min\{1/2, a|\eta|\})$. Finally, to get the explicit form of $\eta$, first note that from \eqref{eq_linear_system} it holds that $h_2^* = -h_1^* w_s/w_d$, so $\eta = h_1^*/(2n_1n_2w_d)$. Then from the expressions of $\delta$ and $\chi_1$ given in \eqref{eq_defn_delta} and \eqref{eq_chi1_chi2} respectively, the conclusion follows. \hfill$\Box$

\bibliographystyle{ieeetr}        % Include this if you use bibtex 
\bibliography{bibliography.bib}           % and a bib file to produce the 
                                 % bibliography (preferred). The
                                 % correct style is generated by
                                 % Elsevier at the time of printing.

\end{document}